\documentclass[utf8]{frontiersinFPHY_FAMS} 

\setcitestyle{square} 
\usepackage{url,hyperref,lineno,microtype,subcaption}
\usepackage[onehalfspacing]{setspace}
\usepackage{parskip}
\usepackage{amssymb,bm}
\usepackage{graphicx}
\usepackage{amsmath}
\usepackage{color}

\def\keyFont{\fontsize{8}{11}\helveticabold }
\def\firstAuthorLast{Yan-Mei Yu {et~al.}} 
\def\Authors{Yan-Mei Yu\,$^{1,4*}$, B. K. Sahoo\,$^{2,*}$, and Bing-Bing Suo\,$^{3}$}
\def\Address{$^{1}$Beijing National Laboratory for Condensed Matter Physics, Institute of Physics, Chinese Academy of Sciences, Beijing 100190,China\\
	$^{2}$Atomic, Molecular and Optical Physics Division, Physical Research Laboratory, Navrangpura, Ahmedabad 380009, India\\
	$^{3}$Institute of Modern Physics, Northwest University, Xi'an, Shaanxi 710069, China\\
	$^{4}$ University of Chinese Academy of Sciences, Beijing 100049, China }
\def\corrAuthor{Yan-Mei Yu}
\def\corrEmail{ymyu@iphy.ac.cn}

\begin{document}
\onecolumn
\firstpage{1}

\title[Highly Charged Ion Clocks]{Highly Charged Ion (HCI) Clocks: Frontier candidates for testing variation of fine-structure constant} 

\author[\firstAuthorLast ]{\Authors} 
\address{} 
\correspondance{} 

\extraAuth{B. K. Sahoo \\ bijaya@prl.res.in}

\maketitle

\begin{abstract}

\section{}

Attempts are made to unify gravity with the other three fundamental forces of nature. As suggested by higher dimensional models, this unification may require space and time variation of some of the dimensionless fundamental constants. In this scenario, probing temporal variation of the electromagnetic fine structure constant ($\alpha= \frac{e^2} {\hbar c}$) in a low energy regime at the cosmological time scale is of immense interest. Atomic clocks are ideal candidates for probing $\alpha$ variation because their transition frequencies are measured with ultra-high precision. Since atomic
transition frequency is a function of $\alpha$, measurement of a clock frequency at different temporal and spatial locations can yield signatures
to ascertain variation of $\alpha$. Highly charged ions (HCIs) are very sensitive to variation of $\alpha$ and are least affected by external perturbations, making them excellent platforms for searching for temporal variation of $\alpha$. In this work, we overview HCIs suitable 
for building atomic clocks because of their spectroscopic features and sensitivity to variation of $\alpha$. The selection of HCI clock candidates
is outlined based on two general rules -- by analyzing trends in fine structure splitting and level-crossing patterns along a series of 
isoelectronic atomic systems of the periodic table. Two variants of relativistic many-body methods in the configuration interaction and 
coupled-cluster theory frameworks are employed to determine the properties of HCIs proposed for atomic clocks. These methods treat electronic 
correlation and relativistic effects under the frame of the Dirac-Coulomb Hamiltonian rigorously, while other higher-order relativistic effects are included 
approximately. Typical systematic effects of the HCI clock frequency measurements are discussed using the calculated atomic properties. This review
will help understand limits and potentials of the proposed HCIs as the prospective atomic clock candidates and guide future HCI clock experiments. 

\tiny
 \keyFont{ \section{Keywords:}Highly charge ions, Optical clocks, fine structure splitting, Level-crossings, Stark shifts, Zeeman shifts,
 	Relativistic many-body methods, Fundamental constants} 
\end{abstract}

\section{Introduction}

Atomic clocks, used for frequency standards, help us define the unit of time with very high precision so that they lose only one second over the
age of our universe. These clocks are based on either neutral atoms or singly charged atomic ions. With the advent of recent laser cooling and 
trapping techniques of atomic systems, modern optical clock frequencies have been measured with uncertainties that are much lower than the present
$10^{-16}$ level caesium microwave-based primary atomic clock \cite{Lodewyck2016,Riehle2021,Boulder2021}. Uncertainty of optical lattice clock 
based on $^{171}$Yb atoms has reached an uncertainty of $1.4\times10^{-18}$ \cite{Schioppo2016,McGrew2018} whereas $^{87}$Sr optical lattice 
clocks offer uncertainty of around $2.0\times10^{-18}$ \cite{Takamoto2016,Oelker2019}. Similarly, uncertainties of $^{171}$Yb$^+$ and 
$^{40}$Ca$^+$ ion clocks have both reached an accuracy of $3\times 10^{-18}$ \cite{Huntermann2016,Huang2022}. There have been continuous efforts 
to reduce uncertainties in the clock frequency measurements and miniaturize the atomic clocks, intending to utilize them in many sophisticated 
instruments and space science research.

Among various scientific applications of atomic clocks, see the review by Safronova et al \cite{Safronova2018} for more details, atomic clocks 
serve as an important tool to probe temporal and spatial variation of the fine-structure constant ($\alpha$). Atomic energy levels are functions 
of $\alpha$. Thus any variation in the $\alpha$ value will result in changes in the energy levels over time and space. Since optical clocks can 
measure atomic transition frequencies to ultra-high precision, they are the most suitable instruments for detecting any drift in the $\alpha$ 
value. The clock transitions have different sensitivity to variation of $\alpha$. The sensitivity of an energy level to $\alpha$ variation is 
gauged through a relativistic sensitivity coefficient $q$ by defining it as
\begin{equation}
	\omega_t=\omega_0+qx,
\end{equation}
where $\omega_0$ is the angular frequency of the transition for the present-day value of the fine-structure constant $\alpha(0)$ and $\omega_t$ is
the angular frequency of the transition corresponding to another value of $\alpha(t)$ at time $t$ such that $x=(\alpha(t)/\alpha(0))^2-1 
\approx 2(\alpha(t)-\alpha(0))/\alpha(0)$. Sometimes it is convenient to introduce a dimensionless sensitivity coefficient $K_{\alpha}$ by defining
it as $K_{\alpha}=q/\omega_0$. Here, $q$ and $K_{\alpha}$ are often referred to as a frequency's absolute and relative sensitivity factor to 
variation of $\alpha$. 

\begin{table}[t]
\centering
\caption{A summary of the sensitivity coefficients, $q$ and $K_{\alpha}$, of the clock transition frequency to variation of $\alpha$, with definition given in text and the data source is taken from Ref. \cite{Flambaum2009}} 
{\setlength{\tabcolsep}{8pt}
\begin{tabular}{c lllll}\hline\hline
$Z$	&	atoms	&	Clock transition 	&	Energy (cm$^{-1}$)	&	$q$  (cm$^{-1}$)  	&	$K_{\alpha}$ (unit less)	\\ \hline
13	&	Al$^+$	&	$3s^2~^1S_0-3s3p~^3P^{\circ}_0$	&	37393.03 	&	146	&	0.008 	\\ [+2ex]
20	&	Ca$^+$	&	$4s~^2S_{1/2}-3d~^2D_{5/2}$	&	13710.89 	&	1040	&	0.152 	\\ [+2ex]
38	&	Sr	&	$5s^2~^1S_0-5s5p~^3P^{\circ}_0$	&	14317.52 	&	443	&	0.062 	\\ [+2ex]
70	&	Yb	&	$6s^2~^1S_0-6s6p~^3P^{\circ}_0$	&	17288.44 	&	2714	&	0.314 	\\ [+2ex]
70	&	Yb$^{+}$	&	$4f^{14}6s~^2S_{1/2}-4f^{14}5d~^2D_{5/2}$	&	24332.69 	&	12582	&	1.03 	\\ [+2ex]
&		&	$4f^{14}6s~^2S_{1/2}-4f^{13}6s~^2F^{\circ}_{7/2}$	&	21418.75 	&	-63752	&	-5.95 	\\ [+2ex]
80	&	Hg	&	$6s^2~^1S_0-6s6p~^3P^{\circ}_0$	&	37645.08 	&	15299	&	0.81 	\\ [+2ex]
80	&	Hg$^{+}$	&	$5d^{10}6s~^2S_{1/2}-5d^{9}6s^2~^2D_{5/2}$	&	35514.62 	&	-52200 	&	-2.94 	\\ \hline\hline
\end{tabular}}	\label{alphasenatom}
\end{table}

Table \ref{alphasenatom} summarizes the currently prevailing clock frequencies with regard to their sensitivity to variation of $\alpha$. Some of 
them has low sensitivities to $\alpha$ variation, like the clock transition in Al$^{+}$, Ca$^{+}$, Sr and Yb, while the 
$4f^{14}6s~^2S_{1/2}-4f^{13}6s~^2F^{\circ}_{7/2}$ transition in Yb$^+$ and the $5d^{10}6s~^2S_{1/2}-5d^{9}6s^2~^2D_{5/2}$ transition in Hg$^{+}$ 
have relatively higher sensitivity coefficients to variation of $\alpha$. Such contrast sensitivities to variation of $\alpha$ suggest different accuracy requirements in the frequency 
measurements if they are used for testing variation of $\alpha$. Assuming the time variation rate of $\alpha$ per year is around $10^{-16}$ level, to detect such a small variation rate of $\alpha$ within a year time, minimum requirements of fractional uncertainties of
the Sr, Yb, and Hg atomic clock frequencies would be about $8\times10^{-17}$, $5\times10^{-17}$ and $2\times10^{-18}$, respectively, by considering their sensitivity coefficients to $\alpha$ variation. Thus, for any further constraining to the temporal variation of $\alpha$ lower than $10^{-16}$, the aimed uncertainty level in the clock frequency measurement has to be less than $10^{-18}$ if an a-year-around experiment is planned. On the one hand, researchers endeavour to push for minimizing uncertainty levels further down in the clock frequency measurements to meet the requirement for inferring any signature of $\alpha$ variation from the measurements, and on the other hand, selecting a suitable pair of clock 
transitions having very different sensitivity to variation of $\alpha$ is a good option. To be specific, when a measurement of the frequency of an
optical clock at the frequency $f_1$ relative to another optical clock at the frequency $f_2$ is conducted (the two optical clocks have different 
sensitivities to $\alpha$ variation), the time-variation of the ratio $f_1/f_2$ can be expressed as 
\begin{equation}
	\frac{\partial}{\partial t}\ln{\frac{f_1}{f_2}}= \frac{\partial}{\partial t}\ln R = \ln \dot{R} =[K_{\alpha}(1)-K_{\alpha}(2)]\frac{\partial}{\partial t}\ln{\alpha} ,
\end{equation}
where $R= \frac{f_1}{f_2}$ is the frequency ratio, $K_{\alpha}(1)$ and $K_{\alpha}(2)$ are the sensitivity factors of the respective clock
frequency. For example, the $^{199}$Hg$^+$ and $^{27}$Al$^{+}$ optical frequency standards have a big difference in $K_{\alpha}$ as 
$K_{\alpha}$(Al$^+$)-$K_{\alpha}$(Hg$^{+}$)=2.95, and their frequency ratio of $f_{\rm{Al}^+}/f_{\rm{Hg}^+}$ has been measured over a timescale of
about one year, yielding a drift rate of $f_{\rm{Al}^+}/f_{\rm{Hg}^+}=(-0.53\pm0.79)\times10^{-16}$ and constrained the temporal variation of
$\alpha$ as $\dot{\alpha}/\alpha=(-1.6\pm2.3)\times10^{-17}$ \cite{Rosenband2008}. Similarly, Lange in PTB compared two optical clocks based on the
E2 and E3 transitions of $^{171}$Yb$^{+}$ that have the difference in $K_{\alpha}$ as $K_{\alpha}$(E3)-$K_{\alpha}$(E2)$=-6.95$. The frequency ratio
$f_{E3}/f_{E2}$ was measured over 1500 days, which determined $\dot{R}/R=-6.8(7.5)\times10^{-18}$ per year and yield 
$\dot{\alpha}/\alpha=(1.0\pm1.1) \times10^{-18}$ per year. 

\begin{table}[t]
	\caption{Time-varying frequency ratio $R$ measurements using different combinations of optical atomic clocks ($A_1$ and $A_2$) , and their corresponding $\alpha$ varying sensitivity coefficients from two separate works. The inferred limits on the time drift of $\alpha$ are also given from these works.}
	{\setlength{\tabcolsep}{12pt}
		\begin{tabular}{lcc cc}\hline\hline
			$A_1-A_2$ &$\dot{R}/R$ & $K_{\alpha}(A_1)-K_{\alpha}(A_2)$	 & $\dot{\alpha}/\alpha$   &Refs. \\ \hline
			Al$^{+}$-Hg$^{+}$& $(-0.53\pm0.79)\times10^{-16}$	&2.95 & $-1.6\pm2.3\times10^{-17}$ &\cite{Rosenband2008} \\
			Yb$^{+}$ E2-E3   & $(-6.8\pm7.5)\times10^{-18}$     & $-6.95$ & $1.0\pm1.1\times10^{-18}$  &\cite{Lange2021}  \\ \hline\hline
	\end{tabular}}
	\label{alphatime}
\end{table} 

In the aim of ultimately observing the variation of $\alpha$, it is necessary to improve the clock frequency measurements consistently. In this sense, even improving the limit on the constraint of $\alpha$ variation over time is meaningful. One way of proceeding with this task is to work hard to find 
ways to minimize uncertainties in the existing atomic clocks. Alternatively, we can look for other suitable candidates least influenced 
by stray electromagnetic fields. One such possibility is to use nuclear transitions for the clock frequency measurements. The presence of a 
low-lying metastable state in $^{229m}$Th with excitation energy around 8 eV is identified to be the perfect choice for this purpose \cite{Peik2009,
	Campbell2012-2}. Many efforts are already gone into directly detecting nuclear transition by measuring the isomeric state energy 
of $^{229m}$Th precisely for achieving nuclear clock \cite{Wense2016,Seiferle2019,Sikorsky2020,Beeks2021}. This transition has also been explored 
for testing variation of fundamental constants by different groups \cite{Flambaum2006,Berengut2009,Fadeev2020,Fadeev2022}. 
The other novel thought has been put into exploring highly charged ions (HCIs) for making atomic clocks \cite{Kozlov2018}. Investigation of spectral properties of HCIs has a very long history, which is immensely interesting for identifying abundant elements in the solar corona and other astrophysical objects, describing exotic phenomena in the 
nuclear reactions, diagnosing plasma processes, etc. \cite{Raymond1989,Gillaspy2001,Prochaska2003,Webb2011}. However,
HCIs are recognized as potential candidates for making ultra-precise atomic clocks, developing tools for quantum information, and probing possible 
variations of $\alpha$. Following the pioneering works by Berengut et al., which suggested enhanced sensitivity to variation of $\alpha$ 
in the HCIs \cite{Berengut2010,Berengut2011,Berengut2011-2,Berengut2012,Berengut2012-3,Berengut2012-4}, a long list of HCIs are being proposed
\cite{Dzuba2012,Dzuba2012-2,Yudin2014,Safronova2014,Safronova2014Ag,Safronova2014Cd,Dzuba2015,Dzuba2015-2,Dzuba2015-3,Yu2016,Yu2018,Yu2019,Porsev2020,
	Beloy2020,Beloy2021,Yu2020,Allehabi2022,Allehabi2022-2} to select appropriate HCIs for consideration for the clock experiments. All these proposed HCI candidates appear  
promising for building atomic clocks and probing $\alpha$ variation, but there could be slight differences in setting up their experiments. Since 
the HCIs possess very compacted electron orbitals than their isoelectronic neutral atoms or singly charged ions, they can be strongly immune to  
external perturbations. Apart from this, it would be relatively easier to detect a signature of $\alpha$ variation in HCIs at the same level of 
accuracy in the clock frequency measurements with respect to optical lattice clocks or singly charged ion clocks owing to their enhanced sensitivity $\alpha$
coefficients.

In this brief review, we shortlist the HCIs that can be prospective candidates for optical atomic clocks other than those previously listed in a 
review by Safronova et al. \cite{Safronova2018}. Most of these HCIs have large $\alpha$ varying sensitivity coefficients for the clock transitions, 
which implies that they are excellent platforms for searching of temporal variation of $\alpha$. Selection of HCIs for atomic clocks is further 
addressed based on the forbidden transitions within fine-structure splitting and those among interconfigurations that undergo orbital energy crossings. 
Relativistic many-body methods employed to calculate HCI properties are described briefly. Finally, prominent systematic 
effects of the HCI clock frequency measurements due to stray electromagnetic fields are discussed before summarizing the work. 

\section{Importance of $\alpha$ -variation studies}

Actual values of many of the dimensionless physical constants need to be verified experimentally as their magnitudes are predicted differently by
different models with strong scientific arguments. Though there has yet to be strong evidence to show that their values are not constant,  
there is also no strong scientific argument justifying that they have to be constants over time or space. Thus, any plausible signature in the temporal 
or spatial variation of fundamental physical constants would answer either of these questions. If slight
temporal or spatial variation in the fundamental constants is possible, it will support some of the phenomena beyond the 
Standard Model of particle physics. This is why probing temporal and spatial variation of $\alpha$ is of immense interest to the research 
community \cite{Uzan2003,Srianand2004,Webb2011,Safronova2018}. One of the biggest consequences of investigating the temporal
and spatial variation of $\alpha$ is to support multidimensional theories that try to unify all four fundamental interactions \cite{Damour1, 
	Damour2, Damour3, Kaluza, Klein, Chodos,Marciano,Bronnikov}. Such theories also predict violation of Einstein's equivalence principle. Thus, any 
signature of variation in the $\alpha$ value can also imply violation of Einstein's equivalence principle. 

In astronomy, observation of absorption spectral lines from distant astronomical objects such as quasars with large red shift constant ($z$) can be 
used directly for probing variation of $\alpha$. Observations using the Kerk telescope suggest that the fractional drift in $\alpha$ is 
smaller by an amount $\Delta \alpha/\alpha$=$(\alpha_z-\alpha_0)/\alpha_0$=$(0.543\pm0.116) \times10^{-5}$ around $10^{10}$ years ago over the range $0.2<z<4.2$ 
\cite{Webb2001,Murphy2003}. Observations of spectral lines from quasar J1120$+$0641 in the redshift range 5.5 to 7.1 using the Very Large 
Telescope (VLT) is used to search for variation of $\alpha$. It is worth mentioning that observations at $z$=7.1 correspond to the Universe 
about 0.8 billion years old. These observations reported the result as $\Delta \alpha/\alpha$=$(-2.18\pm7.27)\times10^{-5}$, which is consistent with null temporal variation \cite{Michael2020}. Very recently,
Murphy et al. observed HE 0515-4414 by the VLT with the redshift range of 0.6-2.4. They combined their observed spectral lines data with 28 
measurements from other spectrographs to mitigate the wavelength calibration errors and reported a weighted mean of $\Delta \alpha / \alpha$=
$(-0.5\pm 0.5_{\text{stat}} \pm 0.4_{\text{sys}}$) parts-per-million (ppm). Webb et al. proposed a theory of spatial dipole of $\alpha$ by analyzing
observed data from both the Keck and VLT telescopes \cite{Webb2011}. It does not seem to have strong evidence in favour of this theory
as data from both the Keck and VLT observations used for this analysis had significant systematic effects \cite{Whitmore2014,Dumont2017}.  

Another class of data is inferred from the geophysical data analysis from the Oklo natural fission reactor and cosmological observations, which 
correspond to a slightly smaller time scale than the astrophysical observations. The signature of temporal variation of $\alpha$ can be easily 
detected from events that correspond to the early epoch of the Universe. The data from the Oklo nuclear reactor offers a limit as
$\Delta \alpha/\alpha<1.1\times10^{-8}$ over a time span of 2 Gyr \cite{Davis2019}.

Laboratory-based atomic clocks provide measurements of the time variation of $\alpha$ in the present-day time scale. Table \ref{alphatime} summarizes
the current most stringent limit to $(\Delta \alpha/ \Delta t)/\alpha= \dot{\alpha}/ \alpha$ per year, which is determined by the laboratory 
measurement of the ratio of two clock frequencies over a reasonably long time. It includes measurements of the clock frequency ratio of the Hg$^+$ 
and Al$^+$ optical clocks over the course of a year ($\Delta t =1$yr), which yields as $\dot{\alpha}/ \alpha=(-1.6 \pm 2.3) \times 10^{-17}$ 
\cite{Rosenband2008}. The other one is the ratio deduced from the electric quadrupole (E2) and electric octupole (E3) clock frequency measurements in Yb$^+$ over around four years ($\Delta t =4$yr), which suggests $\dot{\alpha}/ \alpha=(1.0\pm 1.1 ) \times 10^{-18}$ per year \cite{Lange2021}, agreeing with the previous 
finding of null variation. Similar interpretations on the time variation of $\alpha$ have been made from the other clock frequency measurements 
found from Refs. \cite{Peik2004,Fortier2007,Guena2012,Leefer2013,Godun2014,Huntemann2014,Ashby2018}.

\begin{table}[t]
\caption{Comparison of various limits to variation in $\alpha$ with its absolute value ($\Delta \alpha/\alpha$) obtained from different types of works over a time interval $\Delta t$. The corresponding limits in annual fractional rate of change $(\dot{\alpha}/\alpha)$, with a crude assumption of a linear change in time, are also given for the purpose of comparing them with the values given in Ref. \cite{Lea2007}if one wishes to do so.}
{\setlength{\tabcolsep}{9pt}
\begin{tabular}{lcc cc}\hline\hline
Source & $\Delta\alpha/\alpha$&$\Delta t$ (year)&$\dot{\alpha}/\alpha$(year$^{-1}$)  & Refs. \\ \hline
Cosmological & $<10^{-3}$                  &$10^{10}$   & $10^{-13}$                  &\cite{Uzan2003} \\
Astrophysical(quasar)& $(2.18\pm7.27) \times10^{-5}$  &$10^{10}$   & $(2.18\pm7.27)\times10^{-15}$ &\cite{Michael2020}  \\
Geophysical(Oklo) & $<1.1\times10^{-8}$&$2.1\times10^{9}$&$<10^{-17}$&\cite{Davis2019} \\
Laboratory  &               &     &  $(1.0\pm1.1) \times10^{-18}$        & \cite{Lange2021}     \\\hline\hline
\end{tabular}}
\label{alphatimeall}
\end{table}

Findings of variation of $\Delta \alpha$ from different studies have been reviewed in Refs. \cite{Uzan2003,Karshenboim2000,Lea2007,Flambaum2009,
Berengut2012-3}. Table \ref{alphatimeall} compares limits to the temporal variation in $\alpha$ from various sources. The interpretation of the
astrophysical observations and geophysical data, principally the Oklo phenomenon, are highly dependent on many assumptions. Limits derived from 
the laboratory studies by comparing frequencies of the atomic clocks in that sense can be much less unambiguous. However, these results can only
restricted to the present day phenomena and to the region of space of the earth’s orbit. The time drift of the ratio of two atomic frequencies 
can also depend on the proton-to-electron mass ratio via $\mu_N/\mu_B$=1/$\mu$ when at least one of the atomic clock frequencies is based on the 
hyperfine transition like the Rb and Cs microwave clocks \cite{Guena2012,Godun2014,Huntemann2014,Lange2021}. Together this quantity with $\alpha$ 
variation can tell us about possible variation of a strong-interaction parameter $X_q$, that denotes the ratio between the average quark mass and 
the quantum chromodynamic scale $\Lambda_{QCD}$. Therefore, comparison of two optical clock frequencies is quite important in order to infer 
directly signatures of the variation of the fundamental constants.  

\section{Potential candidates of HCI clocks}

In the past decade, many HCIs have been proposed for making atomic clocks. A list of these ions and their proposed clock transitions
are mentioned in Table \ref{alphasenHCI}. Transition energies and their sensitivity coefficients $q$ and $K$ are also given in the same
table. The first optical HCI clock, based on the Ar$^{13+}$ ions, has been demonstrated at the uncertainty level 2.2$\times10^{-17}$. It used 
sympathetic cooling techniques to decrease temperature to milli-kelvin to cool the ions produced using the electron beam ion 
trap (EBIT). The accuracy of this HCI based optical clock is already comparable to that of many optical clocks in its first attempt itself 
\cite{Micke2020,King2022}. The clock transition in the Ar$^{13+}$ ion was proposed by Yudin et al. \cite{Yudin2014} in 2014, along with many other HCIs having magnetic-dipole (M1) transitions between the 
$np~^2P_{1/2}-^2P_{3/2}$ and among the $np^2~^3P_{0,1}$ fine structure splitting of many monovalent and divalent HCIs respectively. Later, we analyzed major systematics theoretically of clock transitions in the monovalent B-like, Al-like, and Ga-like ions \cite{Yu2016,Yu2019}.


Based on the electric quadrupole (E2) transition between the $np^4~^3P_{0,2}$ fine structure splitting, the Ni$^{12+}$ ion along with the other three HCIs, Cu$^{13+}$, Pd$^{12+}$ and Ag$^{13+}$, were proposed as promising candidates for making high-accuracy optical clocks \cite{Yu2018}. The transition rate of the E2 transition is usually far slower than that of the M1 transition. Therefore the advantage of considering the E2 transitions in these ions for clocks is that they all possess larger quality factors than the M1 transitions between the $(np)~^2P_{1/2,3/2}$ fine structure splittings. The spectroscopy lines of the Ni$^{12+}$ ion in the EBIT has been identified by Liang et al. very recently \cite{Liang2021}. Ba$^{4+}$ with $5s^25p^4$ configuration was proposed by Beloy \cite{Beloy2020} as the mediumly ionized clock.  
The $\alpha$ varying sensitivity coefficients in the Ar$^{13+}$, Ni$^{12+}$ and Ba$^{4+}$ ions are found to be much larger than most of the 
available singly charged ion and neutral atomic clocks except Hg$^+$ and Yb$^+$ clocks. 

Among all the proposed HCIs, Ir$^{17+}$ shows having the largest sensitivity to variation of $\alpha$ \cite{Berengut2011}. The double-hole 
configuration of Ir$^{17+}$ offers several transitions in the optical range. Two of them, $4f^{13}5s~^3F^{\circ}_{4}-4f^{14}~^1S_0$ and 
$4f^{13}5s~^3F^{\circ}_{4}-4f^{12}5s^2~^3H_6$, are claimed to be suitable as clock transitions \cite{Berengut2011}. The optical lines in Ir$^{17+}$ 
along with other Nd-like W, Re, Os, and Pt ions, have been identified by Winderger el al. \cite{Windberger2015}. Their inferred transition energy
of the $4f^{13}5s~^3F^{\circ}_{4}-4f^{12}5s^2~^3H_6$ transition lies between 4.0392 eV and 4.6397 eV. However, this value disagrees
with the theoretical values that predict 3.7623 eV \cite{Cheung2020} and 3.003 eV \cite{Windberger2015}. Similarly, the $4f^{13}5s~^3F^{\circ}_{4}-4f^{14}~^1S_0$ 
transition that has been identified to have the largest $q$ and $K$ values are not observed experimentally, while transition energies from 
different theoretical results show about 20\% difference \cite{Windberger2015,Cheung2020}. The Cf$^{16+}$ and Cf$^{17+}$ ions proposed by Berengut 
et al. \cite{Berengut2012} and the Cf$^{15+}$ ion proposed by Dzuba \cite{Dzuba2015-3} for atomic clocks also have relatively large $q$ and $K$ 
values. Porsev et al. have also explored the possibility of developing optical clocks using the Cf$^{15+}$ and Cf$^{17+}$ ions \cite{Porsev2020}. 

\begin{table}[t]
\centering
\caption{A summary of the HCIs proposed for making optical clocks. The definition of $q$ and $K_{\alpha}$ is given in text.} 
{\setlength{\tabcolsep}{8pt}
\begin{tabular}{c lll lll}\hline\hline
$Z$	&	atoms	&	Clock transition 	&	Energy (cm$^{-1}$)	&	$q$  (cm$^{-1}$)  	&	$K_{\alpha}$ (unit less)	&	Refs. 	\\ \hline
18	&	Ar$^{13+}$	&	$2p_{1/2} - 2p_{3/2}$	&	22656.22	&	22137	&	1.95 	&	\cite{Yudin2014,King2022,Yu2018}	\\ [+2ex]
28	&	Ni$^{12+}$	&	$3s^23p^4~^3P_{2} - ^3P_{0}$	&	20060	&	14982 	&	1.49 	&	\cite{Yu2018,Liang2021}	\\ [+2ex]
56	&	Ba$^{4+}$	&	$5s^25p^4~^3P_2 - ^3P_0$	&	11302	&	5900	&	1.04 	&	\cite{Beloy2020,Allehabi2022}	\\ [+2ex]
82	&	Pb$^{2+}$	&	$6s^2~^1S_0 - 6s6p~^3P_0$	&	60397	&		&		&	\cite{Beloy2021}	\\[+2ex]
77	&	Ir$^{17+}$ 	&	$4f^{13}5s~^3F^{\circ}_{4}-4f^{14}~^1S_0$	&	5055 	&	367161	&	145.3 	&	\cite{Berengut2011}	\\
&		&	$4f^{13}5s~^3F^{\circ}_{4}-4f^{12}5s^2~^3H_6$	&	35285 	&	-385367	&	-21.84 	&		\\ [+2ex]
98	&	Cf$^{15+}$	&	$5f6p^2~^2F^{\circ}_{5/2}$ $-$ $5f^26p~^2H^{\circ}_{9/2}$	&	13303	&	380000	&	57.1 	&	\cite{Dzuba2015-3}	\\
&		&		&	16170	&		&		&	\cite{Porsev2020}	\\ [+2ex]
98	&	Cf$^{16+}$ 	&	$5f6p$ $J$=3 $-$ $6p^2$ $J$=0	&	5267	&	370928	&	140.8 	&	\cite{Berengut2012}	\\ [+2ex]
98	&	Cf$^{17+}$ 	&	$5f_{5/2}-6p_{1/2}$	&	18686	&	449750	&	48.14 	&	\cite{Berengut2012}	\\ 
&		&		&	20611	&		&		&	\cite{Porsev2020}	\\ [+2ex]
60	&	Nd$^{13+}$	&	$5s_{1/2}-4f_{5/2}$	&	58897	&	106000	&	3.60 	&	\cite{Dzuba2012-2}	\\ [+2ex]
62	&	Sm$^{15+}$	&	$4f_{5/2}-5s_{1/2}$	&	55675	&	-136000 	&	-4.89 	&	\cite{Dzuba2012-2}	\\ [+2ex]
59	&	Pr$^{9+}$	&	$5p^2~^3P_0-5p4f~^3G_3$	&	20216	&	42721	&	4.23 	&	\cite{Safronova2014}	\\
&		&		&	2201.36 	&	69918	&	6.33 	&	\cite{Bekker2019}	\\ [+2ex]
60	&	Nd$^{9+}$	&	$(5p^24f )^{\circ}_{5/2}-(5p4f^2 )^{\circ}_{9/2}$	&	20594	&	67136 	&	6.52 	&	\cite{Yu2020}	\\ [+2ex]
67	&	Ho$^{14+}$	&	$4f^65s~^8F_{1/2}-4f^55s^2~^6H^{\circ}_{5/2}$	&	23823	&	-186000	&	7.81 	&	\cite{Dzuba2015}	\\ \hline\hline
\end{tabular}}	\label{alphasenHCI}
\end{table}

The ultra-narrow $5s_{1/2}-4f_{5/2}$ optical transitions in Nd$^{13+}$ and Sm$^{15+}$ were proposed for the construction of the HCI clocks and search 
for variation of the fine-structure constant \cite{Dzuba2012-2}. These lines have relatively large $q$ and $K$ values, but their transition 
wavelengths are about 170 nm and 180 nm, which are out of the accessible range of the currently available lasers. Safronova et al. searched for the 
visible lines in a series of the Ag-like, Cd-like, In-like, and Sn-like HCIs for the development of frequency standards and to probe variation of the 
fine-structure constant \cite{Safronova2014,Safronova2014Ag,Safronova2014Cd}. Among these ten HCIs recommended by them, the Pr$^{9+}$ ion
seems to be the more attractive, which has an optical transition $5p^2~^3P_0-5p4f~^3G_3$ between the ground and the first excited states. The 
lifetime of the excited clock state was estimated to be $10^{14}$ s, and it has a wavelength at 475 nm, which can be accessed by the currently 
available. Bekker et al. present the EBIT measurements of the spectra of Pr$^{9+}$, which refines the energy levels of such proposed nHz-wide 
clock line and demonstrates about Pr$^{9+}$ having very large sensitivity coefficient to variation of $\alpha$ and variation of local Lorentz 
invariance probing coefficient \cite{Bekker2019}. The Sb-like Nd$^{9+}$ ion was proposed by Yu et al. \cite{Yu2020} as a possible HCI clock. The 
Nd$^{9+}$ ion has the ground state configuration as $(5p^24f )^{\circ}_{5/2}$ and the first excited state configuration as  
$(5p4f^2 )^{\circ}_{9/2}$. The E2 transition of $(5p^24f )^{\circ}_{5/2}-(5p4f^2 )^{\circ}_{9/2}$ shows large $q$ and $K$ values and 
feasibility of making a high accuracy atomic clock. 

Dzuba et al. studied the I-like Ho$^{14+}$ ion as a possible candidate for an extremely accurate and stable optical clock which has $q$ and $K$ 
slightly larger than Pr$^{9+}$ and Nd$^{9+}$. The Ho$^{14+}$ ion has all the desired features, including relatively strong optical electric-dipole 
(E1) and M1 transitions can be used for cooling and detection. The electronic configuration of the ground state of this ion has seven valence 
electrons in the open $s$ and $f$ shells, which gives rise to a large number of fine structure splitting. Nakajima et al. have observed the visible spectra 
of the above ion using a compact electron beam ion trap \cite{Nakajima2017}. They found that the ground state configuration of Ho$^{14+}$ is [Kr]$4f^65s$ with 577 fine structure levels spreading over an energy range of about 40 eV, and the electronic configuration of the first-excited configuration of this
ion, [Kr]$4f^55s^2$, with 200 fine structure levels spreading over the similar energy range. Their results show a huge number of emissions in
the Ho$^{14+}$ visible spectra, whereas these lines are challenging to distinguish while experimenting as they all have similar transition probabilities.

\section{General directions for identifying HCI clock candidates}

The HCIs have very compact sizes and wave functions compared with their counter neutral atoms and singly charged ions, which are accompanied with 
the scaling laws for energies and transition matrix elements with increasing the charge number $Z_{\rm{ion}}$ in an isoelectronic sequence 
\cite{Gillaspy2001,Berengut2013}. In order to understand the enhancement of $q$ in the HCIs, Berengut et al suggested a simple analytical expression as \cite{Berengut2013,Berengut2010}  
\begin{equation}
	q_n \approx -I_n \frac{(Z\alpha)^2}{\nu (j+1/2)}	\label{deltan}
\end{equation}
and
\begin{equation}
	I_n =-E_n=\frac{Z^2_a}{2n^2}=\frac{(Z_{ion}+1)^2} {2\nu^2} , \label{In}
\end{equation}
where $Z$ is the atomic number, $j$ is the total angular momentum of the orbital, $I_n=-E_n$ represents the positive ionization potential energy
of the electron in the electronic orbital, and $n$ is the integer principal quantum number. The equality in Eq. (\ref{In}) relates $I_n$ to $E_n$ by 
using an effective principle quantum number $\nu$ and an effective charge $Z_a$ that the electron sees at larger distances, considering the 
screening in a multi-electron system. Eqs. (\ref{deltan}) and (\ref{In}) explain how enhancement of $q$ increases with $I_n$, i.e., with larger 
$Z_{ion}$. Besides, $q$ is related to $j$, which indicates that the excitation among orbitals with larger $j$ differences can have bigger $q$. Appearance of
$Z$ means a heavy element is preferred, usually offering larger relativistic effects. Besides, the hole state in an otherwise filled external shell
will have the highest $q$ for particular values of $Z$ and $Z_a$.

The energy scaling with $Z_{\rm{ion}}$ indicates that transitions in HCIs are frequently in the XUV and x-ray range unsuitable for making 
atomic clocks. However, we can realize that on occasions where HCIs have forbidden transitions can be in the optical region, as has been pointed out 
by the previous studies. Firstly, the energy interval between the fine structure and the hyperfine structure energy levels increase greatly 
with the charge number $Z_{\rm{ion}}$ in the HCI isoelectronic sequences, which leads to widening these level splitting. It causes transition lines 
to change from the microwave, as seen in the neutral atoms and singly charged ions, to the optical in HCIs. The other possible approach can be 
adopted to find suitable HCIs for clocks by analyzing the orbital energy crossings in the isoelectronic sequences of atomic systems. Near level 
crossings, frequencies of transitions involving the orbital sequence switching can be much smaller than the ionization potentials. This helps to 
distinguish suitable transitions that can be in the optical range in a prudent manner.

In the following, we overview two basic rules for selecting HCIs as candidates for making atomic clocks. In fact, from a general observation point 
of view of all the
previously proposed HCI candidates for clocks, can be classified into two categories: HCIs with the M1 and E2 clock transitions among the fine 
structure splitting and HCIs with the forbidden transitions among the electronic configurations whose candidatures can be understood using orbital 
energy crossing scheme. Thus, we proceed further to investigate more HCIs along the same lines of thoughts to search for other suitable candidates for 
HCI clocks. 

\subsection{Forbidden transitions within fine-structure splitting}

Studies of spectroscopic properties of forbidden lines are attractive for quite a long time in the fields of astronomical and laboratory research since they are frequently observed in astrophysical observations, e.g. in the solar corona and in solar flares, and high-temperature fusion plasma, where these forbidden lines are often used as diagnostic tools to find out temperature, constituents and ion densities \cite{Davidson1979,Tsamis2003,
	Trabert2000,Trabert2014,Brewer2016,Wahlgren2021}. The M1 transition lines occur amongst the fine-structure splitting of the HCI isoelectronic sequences and are usually found in the optical range. They are easier to observe in contrast to the resonant lines that appear in the far ultraviolet region. Biemont and coworkers have carried out comprehensive studies on many radiative properties of transition lines in HCIs having ground state configurations as $3p^k$, $4p^k$, $5p^k$, and $6p^k$ \cite{Biemont1983, Biemont1986,Biemont1987,Biemont1988,Biemont1996,Sugar1990,Shirai1991,
	Chou1996,Bhatia1998,Trabert2004,Trabert2008} with $k=1,2,3,\cdots$ number of valence electrons. 

Recently, many optical forbidden transitions between the fine-structure splitting in the HCIs are considered as clock transitions, and many of them are listed in Table \ref{M1E2Fine}. As can be seen from the above table, many 
proposals have considered the forbidden M1 transitions between the fine-structure splitting $^2P_{1/2}-^2P_{3/2}$ in HCIs having single $p$-valence orbital and between the fine -structure splitting $^3P_0-^3P_1$ in HCIs having two $p$ valence electrons for making ultra-precise atomic clocks. Yu et al. have investigated B-, Al-, and Ga-like HCIs for making atomic clocks that have fine-structure splitting as $2p_{1/2,3/2}$, $3p_{1/2,3/2}$ 
and $4p_{1/2,3/2}$ respectively. It was found in this work that the M1 transitions between these fine structure splitting have appropriate wavelengths and quality factors as well as relatively smaller systematics (at the 10$^{-19}$ level). Therefore, these ions are appeared to be excellent candidates to be considered for making atomic clocks \cite{Yu2016,Yu2018}. One can presume from these analyses that 
HCIs with $np^k$ ground state configurations with $n\geq 5$ can have similar transition properties and be suitable for atomic clocks. 

Compared with M1 transitions, wherever E2 transitions play important roles in deciding the lifetimes of an upper state, they can offer larger quality 
factors if considered as clock transitions. Yu et al. have proposed the S-like Ni$^{12+}$ and Cu$^{13+}$ ions having the $3p^4$ ground-state 
configurations, and the Se-like Pd$^{12+}$ and Ag$^{13+}$ ions having the $4p^4$ ground-state configurations as potential candidates for making 
an ultra-precise atomic clock, as listed in Table \ref{M1E2Fine}. The $^3P_2-^3P_0$ transitions in the $^{58}$Ni$^{12+}$ and $^{106}$Pd$^{12+}$ ions 
with nuclear spin $I=0$ and the $|^3P_2,~F=1/2 \rangle-|^3P_0,~F=3/2\rangle$ transition in $^{63}$Cu$^{13+}$ with $I=3/2$, and the $|^3P_2,~F=3/2
\rangle-|^3P_0,~F=1/2\rangle$ transition in $^{107}$Ag$^{13+}$ with $I=1/2$ can also be suitable for clock transitions, which can offer 
clock transitions with fractional uncertainties below $10^{-19}$ level. These HCIs have simple atomic energy level structures to carry out 
spectroscopic measurements, and the quality factors of the respective clock transitions are of the order $10^{15}-10^{16}$. Allehabi et al. have 
identified $^3P_2-~^3P_0$ transitions from HCIs of group-16 elements that are governed by E2 channel and are in the optical region as possible 
high-accuracy optical clocks, including the S-, Se-, and Te-like systems, which have $3p^4$, $4p^4$, and $5p^4$ ground configurations 
\cite{Allehabi2022} as listed in Table \ref{M1E2Fine}. 

The forbidden transitions among the fine-structure splitting in HCIs possessing $nd^k$ configuration in their ground state are more complicated to
study than the HCIs having $np^k$ configurations in the ground state. The reason is $d$-orbitals have many subshells bestowing strong open-shell
problems. As a result, it is challenging to study transition properties among the fine-structure splitting of these HCIs, even using the 
available sophisticated atomic many-body methods. Some of these transitions are found to be useful in the production processes 13.5-nm-wavelength 
extreme ultraviolet (EUV) radiation for nanolithographic applications \cite{Torretti2017}. They also give rise EUV spectra between 10 and 25 nm in the 
tungsten HCIs from the $3d$ open-shell configurations as listed in the NIST database \cite{Ralchenko2011}. It is natural to anticipate that some of 
these transitions could be considered for atomic clocks. Theoretical studies carried out in Refs. \cite{Safronova2018-2,Si2018,Biemont1989,Ali1988,
	Charro2002,Trabert2001,Guise2014} support this idea. A wide range of calculations on the M1 and E2 transition probabilities among the fine-structure
splitting of the $3d~^2D_{3/2}–3d~^2D_{5/2}$ transitions are performed in the K-like sequence HCIs to find out their feasibility for using optical 
clocks \cite{Jian2017}. Some of them are listed in Table \ref{M1E2Fine}.

Now we turn to the transitions involving fine-structure partners of open $4f$ shells. For the reason mentioned above in the case of the $nd^k$ 
configurations, theoretical studies of these transitions are even more complicated than the open $p$- and $d$-shells. Electrons from the open-shell 
$4f$ orbitals usually show strong correlation effects with the inner $s$, $p$, and $d$ shells. Historically, the forbidden transitions among the 
fine structures of the open $f$-shell configurations are widely investigated for various applications such as carrying out temperature and 
density diagnostics of plasma, and astronomical spectral identifications \cite{Biemont2003,Radziute2015,Clementson2010,Priti2020,Radziute2021}. A 
series of HCIs from Hf$^{12+}$ to U$^{34+}$ that have $4f^{12}$ valence shells were proposed to be excellent candidates for building exceptionally
accurate atomic clocks by Derevianko et al.\cite{Derevianko2012,Dzuba2012}. Some of them are listed in Table \ref{M1E2Fine}. In addition to the 
highly forbidden laser-accessible transitions within the $4f^{12}$ ground-state configurations that can be used to be the clock transitions, these 
ions also have additional M1 and E2 transitions along with the M1 clock transitions that can be used for cooling ions during the measurement.

In Table \ref{M1E2Fine}, we compare the $q$ and $K$ values for the proposed candidates of the HCI clocks when they are reported earlier. Most of 
them have larger $q$ and $K$ values than their isoelectronic neutral atoms and singly charged ions, however, are still less than the singly charged 
ions Hg$^+$ and Yb$^+$. It indicates that $q$ and $K$ are not significantly prominent in the M1 and E2 transitions between the fine-structure splittings.

\begin{table*}[t]
\scriptsize
\centering
\caption{Summary of selected M1 and E2 HCI transitions of fine-structure splittings. $a[b]$ should be read as $a \times10^{b}$.}
{\setlength{\tabcolsep}{8pt}
\begin{tabular}{c ll ll ll ll ll  }\hline\hline
Ions	&	$Z$	&	Conf.	&	Clock transition	&	Type 	&	$\lambda$ 	&	$\Gamma$  	&	$Q$	&	$q$	&	$K$	&	Refs. 	\\
&		&		&		&		&	(nm)	&	(Hz)	&		&	(cm$^{-1}$)	&		&		\\ \hline
S$^{11+}$	&	16	&	$2p$	&	 $^2P^{\circ}_{1/2}-^2P^{\circ}_{3/2}$	&	M1	&	760 	&	3.27 	&	1.2[14]	&	12795	&	1.94 	&	\cite{Yudin2014,Yu2019}	\\
Cl$^{12+}$	&	17	&	$2p$	&	 $^2P^{\circ}_{1/2}-^2P^{\circ}_{3/2}$	&	M1	&	574 	&	7.59 	&	6.9[13]	&	16960	&	1.95 	&	\cite{Yudin2014,Yu2019}	\\
Ar$^{13+}$	&	18	&	$2p$	&	 $^2P^{\circ}_{1/2}-^2P^{\circ}_{3/2}$	&	M1	&	441 	&	16.71 	&	4.1[13]	&	22137	&	1.95 	&	\cite{Yudin2014,Yu2019}	\\
K$^{14+}$	&	19	&	$2p$	&	 $^2P^{\circ}_{1/2}-^2P^{\circ}_{3/2}$	&	M1	&	345 	&	35.07 	&	2.5[13]	&	28137	&	1.94 	&	\cite{Yudin2014,Yu2019}	\\
Cr$^{11+}$	&	24	&	$3p$	&	 $^2P^{\circ}_{1/2}-^2P^{\circ}_{3/2}$	&	M1	&	816 	&	2.66 	&	1.4[14]	&	-12352	&	-2.01 	&	\cite{Yudin2014,Yu2016}	\\
Mn$^{12+}$	&	25	&	$3p$	&	 $^2P^{\circ}_{1/2}-^2P^{\circ}_{3/2}$	&	M1	&	654 	&	5.12 	&	9.0[13]	&	-15440	&	-2.02 	&	\cite{Yudin2014,Yu2016}	\\
Co$^{14+}$	&	27	&	$3p$	&	 $^2P^{\circ}_{1/2}-^2P^{\circ}_{3/2}$	&	M1	&	436 	&	17.28 	&	4.0[13]	&	-23301	&	-2.03 	&	\cite{Yudin2014,Yu2016}	\\
Cu$^{16+}$	&	29	&	$3p$	&	 $^2P^{\circ}_{1/2}-^2P^{\circ}_{3/2}$	&	M1	&	301 	&	51.95 	&	1.9[13]	&	33836	&	2.04 	&	\cite{Yudin2014,Yu2016}	\\
Nb$^{10+}$	&	41	&	$4p$	&	 $^2P^{\circ}_{1/2}-^2P^{\circ}_{3/2}$	&	M1	&	418 	&	19.76 	&	3.6[13]	&	23928	&	2.00 	&	\cite{Yudin2014,Yu2019}	\\
Ru$^{13+}$	&	44	&	$4p$	&	 $^2P^{\circ}_{1/2}-^2P^{\circ}_{3/2}$	&	M1	&	255 	&	87.01 	&	1.4[13]	&	41741	&	2.13 	&	\cite{Yudin2014,Yu2019}	\\
Xe$^{5+}$	&	54	&	$5p$	&	 $^2P^{\circ}_{1/2}-^2P^{\circ}_{3/2}$	&	M1	&	641 	&	5.43 	&	8.6[13]	&		&		&	\cite{Yudin2014}	\\
Ba$^{7+}$	&	56	&	$5p$	&	 $^2P^{\circ}_{1/2}-^2P^{\circ}_{3/2}$	&	M1	&	424 	&	18.80 	&	3.8[13]	&		&		&	\cite{Yudin2014}	\\
Ni$^{14+}$	&	28	&	$2p^2$	&	 $^3P_{0}-^3P_{1}$	&	M1	&	670 	&	9.5	&	4.7[13]	&		&		&	\cite{Yudin2014}	\\
Br$^{3+}$	&	35	&	$3p^2$	&	 $^3P_{0}-^3P_{1}$	&	M1	&	3814 	&	0.05	&	1.6[15]	&		&		&	\cite{Yudin2014}	\\
Rb$^{5+}$	&	37	&	$4p^2$	&	 $^3P_{0}-^3P_{1}$	&	M1	&	1946 	&	0.39	&	4.0[14]	&		&		&	\cite{Yudin2014}	\\
Xe$^{4+}$	&	54	&	$5p^2$	&	 $^3P_{0}-^3P_{1}$	&	M1	&	1076 	&	2.3	&	1.2[14]	&		&		&	\cite{Yudin2014}	\\
Ba$^{6+}$	&	56	&	$5p^2$	&	 $^3P_{0}-^3P_{1}$	&	M1	&	645 	&	10.7	&	4.3[13]	&		&		&	\cite{Yudin2014}	\\
Ni$^{12+}$	&	28	&	$3p^4$	&	 $^3P_{2}-^3P_{0}$	&	E2	&	498 	&	0.008 	&	7.2[16]	&	22473	&	2.3	&	\cite{Yu2018,Liang2021}	\\
Cu$^{13+}$	&	29	&	$3p^4$	&	 $^3P_{2}-^3P_{0}$	&	E2	&	431 	&	0.011 	&	6.6[16]	&		&		&	\cite{Yu2018}	\\
Pd$^{12+}$	&	46	&	$4p^4$	&	 $^3P_{2}-^3P_{0}$	&	E2	&	396 	&	0.080 	&	9.5[15]	&		&		&	\cite{Yu2018}	\\
Ag$^{13+}$	&	47	&	$4p^4$	&	 $^3P_{2}-^3P_{0}$	&	E2	&	367 	&	0.100 	&	8.2[15]	&		&		&	\cite{Yu2018}	\\
Xe$^{2+}$	&	54	&	$5p^4$	&	 $^3P_{2}-^3P_{0}$	&	E2	&	1230 	&	0.006 	&	3.8[16]	&	5611	&	1.38	&	\cite{Allehabi2022}	\\
Ba$^{4+}$	&	56	&	$5p^4$	&	 $^3P_{2}-^3P_{0}$	&	E2	&	885 	&	0.018 	&	1.9[16]	&	5976	&	1.06	&	\cite{Allehabi2022}	\\
Ce$^{6+}$	&	58	&	$5p^4$	&	 $^3P_{2}-^3P_{0}$	&	E2	&	704 	&	0.034 	&	1.2[16]	&	5907	&	0.83	&	\cite{Allehabi2022}	\\
Zr$^{6+}$	&	40	&	$4p^4$	&	 $^3P_{2}-^3P_{0}$	&	E2	&	816 	&	0.007 	&	5.2[16]	&	8939	&	1.42	&	\cite{Allehabi2022}	\\
Cd$^{14+}$	&	48	&	$4p^4$	&	 $^3P_{2}-^3P_{0}$	&	E2	&	347 	&	0.12 	&	7.1[15]	&	8837	&	0.61	&	\cite{Allehabi2022}	\\
Ge$^{16+}$	&	32	&	$3p^4$	&	 $^3P_{2}-^3P_{0}$	&	E2	&	300 	&	0.038 	&	2.6[16]	&	18484	&	1.11	&	\cite{Allehabi2022}	\\
Kr$^{20+}$	&	36	&	$3p^4$	&	 $^3P_{2}-^3P_{0}$	&	E2	&	213 	&	0.13 	&	1.1[16]	&	17252	&	0.74	&	\cite{Allehabi2022}	\\
Sr$^{22+}$	&	38	&	$3p^4$	&	 $^3P_{2}-^3P_{0}$	&	E2	&	187 	&	0.20 	&	8.1[15]	&	14130	&	0.53	&	\cite{Allehabi2022}	\\
Cu$^{10+}$	&	29	&	$3d$	&	$^2D_{3/2}-^2D_{5/2}$	&	M1	&	2465 	&	0.114 	&	1.1[15]	&		&		&	\cite{Jian2017}	\\
Ga$^{12+}$	&	31	&	$3d$	&	$^2D_{3/2}-^2D_{5/2}$	&	M1	&	1585 	&	0.430 	&	4.4[14]	&		&		&	\cite{Jian2017}	\\
As$^{14+}$	&	33	&	$3d$	&	$^2D_{3/2}-^2D_{5/2}$	&	M1	&	1071 	&	1.394 	&	2.0[14]	&		&		&	\cite{Jian2017}	\\
Se$^{15+}$	&	34	&	$3d$	&	$^2D_{3/2}-^2D_{5/2}$	&	M1	&	893 	&	2.403 	&	1.4[14]	&		&		&	\cite{Jian2017}	\\
Hf$^{12+}$	&	72	&	$4f^{12}$	&	$^3H_6-^3F_4$	&	E2	&	1168 	&	9.5[-6]	&	2.7[19]	&		&		&	\cite{Dzuba2012}	\\
Os$^{18+}$	&	76	&	$4f^{12}$	&	$^3H_6-^3F_4$	&	E2	&	1008 	&	1.4[-5]	&	2.2[19]	&		&		&	\cite{Dzuba2012}	\\
Hg$^{22+}$	&	80	&	$4f^{12}$	&	$^3H_6-^3F_4$	&	E2	&	922 	&	1.3[-5]	&	2.4[19]	&		&		&	\cite{Dzuba2012}	\\
Po$^{26+}$	&	84	&	$4f^{12}$	&	$^3H_6-^3F_4$	&	E2	&	860 	&	1.2[-5]	&	2.9[19]	&		&		&	\cite{Dzuba2012}	\\
Ra$^{30+}$	&	88	&	$4f^{12}$	&	$^3H_6-^3F_4$	&	E2	&	814 	&	1.1[-5]	&	3.2[19]	&		&		&	\cite{Dzuba2012}	\\
U$^{34+}$	&	92	&	$4f^{12}$	&	$^3H_6-^3F_4$	&	E2	&	778 	&	1.1[-5]	&	3.8[19]	&		&		&	\cite{Dzuba2012}	\\ \hline\hline
\end{tabular}}
\label{M1E2Fine}
\end{table*}

\subsection{Orbital energy crossings in complex HCIs}
Generally, the ordering of energy levels in neutral atoms follows the $n+l$ and $n$ orbital filling scheme (often referred to as Madelung rule 
\cite{Madelung1936}), where $n$ and $l$ are the principles and angular quantum numbers. According to this rule, we can assume that the order of
the atomic orbital filling in the neutral and singly charged ions would be $1s \rightarrow 2s \rightarrow 2p 
\rightarrow 3s \rightarrow 3p \rightarrow 4s \rightarrow 3d \rightarrow 4p \rightarrow 5s \rightarrow 4d \rightarrow 5p \rightarrow 6s \rightarrow 
4f \rightarrow 5d \rightarrow 6p \rightarrow 7s \rightarrow 5f \rightarrow 6d \rightarrow 7p \cdots$. However, this rule is violated in the hydrogen-like ions, in which the filling order of the orbitals complies with the Coulombic ordering. That is, orbitals with smaller $n$ are filled first, and for equal $n$, the orbitals with smaller $l$ are filled. If we follow this ordering, then $nd$ orbitals are filled first instead of $(n+1)s$ orbitals after the $np$ orbitals. Similarly, $nf$ orbitals are filled first instead of $(n+1)s$ orbitals after the $nd$. The phenomenon of the orbital energy of $nd$ and $nf$ diving down while $(n+1)s$ and also $(n+1)p$ rising up as the $Z$ and $Z_{ion}$ increasing in an isoelectronic sequence is referred to as the orbital energy crossing, which is first investigated by Berengut, et al., \cite{Berengut2012-4}. They exterminated all the energy crossing in the periodical table. The previously proposed HCIs for making atomic clocks are mostly concentrated on the $nd-(n+1)s$ crossing and the $4f-5s$ crossing.  

\begin{table*}[t]
\scriptsize
\caption{The energy levels of some ions in the K-like ($N=19$), Ca-like ($N=20$), and the forth row transition metals from Sc- to Ni-like ($N=21-28$) isoelectronic sequence HCIs, where $N$ is the number of the electron, $Z$ is the atomic number, and $x$ is the number of valence electrons. The closed-shell configuration is [Ar]$\equiv 1s^2 2s^2 2p^6 3s^2 3p^6$. The data are taken from the NIST database \cite{nist}, and units are in cm$^{-1}$. }	
{\setlength{\tabcolsep}{5pt}
\begin{tabular}{lll lll l | lll lll l}\hline\hline
			$N$  	&	 $Z $	&	$x $	&	Ion      	&	Configuration 	&	State      	&	Energy	&	 $N$  	&	 $Z $	&	$x $	&	Ion      	&	Configuration 	&	State      	&	Energy	\\ \hline
			\multicolumn{7}{c}{K-like isoelectronic sequences}													&	\multicolumn{7}{c}{Ca-like isoelectronic sequences} 													\\
			19	&	20	&	1	&	Ca$^+$   	&	[Ar]$4s$      	&	$^2S_{1/2}$	&	0	&	20	&	21	&	2	&	Sc$^{+}$ 	&	[Ar]$3d4s$    	&	$^3D_{1}$  	&	0	\\
			&	   	&	  	&	         	&	[Ar]$3d$      	&	$^2D_{3/2}$	&	13650	&	    	&	   	&	  	&	         	&	[Ar]$3d^2$    	&	$^3F_{2}$  	&	4803	\\
			&	21	&	1	&	Sc$^{2+}$	&	[Ar]$3d$      	&	$^2D_{3/2}$	&	0	&	    	&	22	&	2	&	Ti$^{2+}$	&	[Ar]$3d^2$    	&	$^3F_{2}$  	&	0	\\
			&	   	&	  	&	         	&	[Ar]$4s$      	&	$^2S_{1/2}$	&	25539	&	    	&	   	&	  	&	         	&	[Ar]$3d4s$    	&	$^3D_{1}$  	&	38064	\\
			&	22	&	1	&	Ti$^{3+}$	&	[Ar]$3d$      	&	$^2D_{3/2}$	&	0	&	    	&	23	&	2	&	V$^{3+}$ 	&	[Ar]$3d^2$    	&	$^3F_{2}$  	&	0	\\
			&	   	&	  	&	         	&	[Ar]$4s$      	&	$^2S_{1/2}$	&	80389	&	    	&	   	&	  	&	         	&	[Ar]$3d4s$    	&	$^3D_{1}$  	&	96196	\\ \hline
			\multicolumn{7}{c}{Sc-like isoelectronic sequences}													&	\multicolumn{7}{c}{Ti-like isoelectronic sequences}													\\
			21	&	22	&	3	&	Ti$^{+}$ 	&	[Ar]$3d^2 4s$ 	&	$^4F_{3/2}$	&	0	&	22	&	23	&	4	&	V$^{+}$  	&	[Ar]$3d^4 $   	&	$^5D_{0}$  	&	0	\\
			&	   	&	  	&	         	&	[Ar]$3d^3 $   	&	$^4F_{3/2}$	&	908	&	    	&	   	&	  	&	         	&	[Ar]$3d^3 4s$ 	&	$^5F_{1}$  	&	2604	\\
			&	23	&	3	&	V$^{2+}$ 	&	[Ar]$3d^3 $   	&	$^4F_{3/2}$	&	0	&	    	&	24	&	4	&	Cr$^{2+}$	&	[Ar]$3d^4$    	&	$^4F_{3/2}$	&	0	\\
			&	   	&	  	&	         	&	[Ar]$3d^2 4s$ 	&	$^4F_{3/2}$	&	43942	&	    	&	   	&	  	&	         	&	[Ar]$3d^3 4s$ 	&	$^5F_{1}$  	&	49492	\\
			&	24	&	3	&	Cr$^{3+}$	&	[Ar]$3d^3 $   	&	$^4F_{3/2}$	&	0	&	    	&	25	&	4	&	Mn$^{3+}$	&	[Ar]$3d^4$    	&	$^4F_{3/2}$	&	0	\\
			&	   	&	  	&	         	&	[Ar]$3d^2 4s$ 	&	$^4F_{3/2}$	&	103996	&	    	&	   	&	  	&	         	&	[Ar]$3d^3 4s$ 	&	$^5F_{1}$  	&	111506	\\ \hline
			\multicolumn{7}{c}{V-like isoelectronic sequences}													&	\multicolumn{7}{c}{Cr-like isoelectronic sequences} 													\\
			23	&	24	&	5	&	Cr$^{+}$ 	&	[Ar]$3d^5$    	&	$^6S_{5/2}$	&	0	&	24	&	25	&	6	&	Mn$^{+}$ 	&	[Ar]$3d^5 4s$ 	&	$^7S_{3}$  	&	0	\\
			&	   	&	  	&	         	&	[Ar]$3d^4 4s$ 	&	$^6D_{1/2}$	&	11961	&	    	&	   	&	  	&	         	&	[Ar]$3d^6   $ 	&	$^5D_{4}$  	&	14326	\\
			&	25	&	5	&	Mn$^{2+}$	&	[Ar]$3d^5$    	&	$^6S_{5/2}$	&	0	&	    	&	26	&	6	&	Fe$^{2+}$	&	[Ar]$3d^6 $   	&	$^5D_{4}$  	&	0	\\
			&	   	&	  	&	         	&	[Ar]$3d^4 4s$ 	&	$^6D_{1/2}$	&	62457	&	    	&	   	&	  	&	         	&	[Ar]$3d^5 4s$ 	&	$^5S_{2}$  	&	41000	\\
			&	26	&	5	&	Fe$^{3+}$	&	[Ar]$3d^5$    	&	$^6S_{5/2}$	&	0	&	    	&	27	&	6	&	Co$^{3+}$	&	[Ar]$3d^6 $   	&	$^5D_{4}$  	&	0	\\
			&	   	&	  	&	         	&	[Ar]$3d^4 4s$ 	&	$^6D_{1/2}$	&	127766	&	    	&	   	&	  	&	         	&	[Ar]$3d^5 4s$ 	&	$^5S_{2}$  	&	102774	\\ \hline
			\multicolumn{7}{c}{Mn-like isoelectronic sequences}													&	\multicolumn{7}{c}{Fe-like isoelectronic sequences}													\\
			25	&	26	&	7	&	Fe$^{+}$ 	&	[Ar]$3d^6 4s$ 	&	$^6D_{9/2}$	&	0	&	26	&	27	&	8	&	Co$^{+}$ 	&	[Ar]$3d^8$    	&	$^3F_{4}$  	&	0	\\
			&	   	&	  	&	         	&	[Ar]$3d^7$    	&	$^4F_{9/2}$	&	1872	&	    	&	   	&	  	&	         	&	[Ar]$3d^7 4s$ 	&	$^3F_{4}$  	&	9814	\\
			&	27	&	7	&	Co$^{2+}$	&	[Ar]$3d^7$    	&	$^4F_{9/2}$	&	0	&	    	&	28	&	8	&	Ni$^{2+}$	&	[Ar]$3d^8$    	&	$^3F_{4}$  	&	0	\\
			&	   	&	  	&	         	&	[Ar]$3d^6 4s$ 	&	$^4D_{7/2}$	&	55729	&	    	&	   	&	  	&	         	&	[Ar]$3d^7 4s$ 	&	$^5F_{4}$  	&	61338	\\
			&	28	&	7	&	Ni$^{3+}$	&	[Ar]$3d^7$    	&	$^4F_{9/2}$	&	0	&	    	&	29	&	8	&	Cu$^{3+}$	&	[Ar]$3d^8$    	&	$^3F_{4}$  	&	0	\\
			&	   	&	  	&	         	&	[Ar]$3d^6 4s$ 	&	$^4D_{7/2}$	&	120909	&	    	&	   	&	  	&	         	&	[Ar]$3d^7 4s$ 	&	$^5F_{4}$  	&	128343	\\ \hline
			\multicolumn{7}{c}{Co-like isoelectronic sequences}													&	\multicolumn{7}{c}{Ni-like isoelectronic sequences} 													\\
			27	&	28	&	9	&	Ni$^{+}$ 	&	[Ar]$3d^9$    	&	$^2D_{5/2}$	&	0	&	28	&	29	&	10	&	Cu$^{+}$ 	&	[Ar]$3d^{10}$ 	&	$^1S_{0}$  	&	0	\\
			&	   	&	  	&	         	&	[Ar]$3d^8 4s$ 	&	$^2F_{7/2}$	&	13550	&	    	&	   	&	  	&	         	&	[Ar]$3d^9 4s$ 	&	$^1D_{2}$  	&	26264	\\
			&	29	&	9	&	Cu$^{2+}$	&	[Ar]$3d^9$    	&	$^2D_{5/2}$	&	0	&	    	&	30	&	10	&	Zn$^{2+}$	&	[Ar]$3d^{10}$ 	&	$^1S_{0}$  	&	0	\\
			&	   	&	  	&	         	&	[Ar]$3d^8 4s$ 	&	$^2F_{7/2}$	&	67017	&	    	&	   	&	  	&	         	&	[Ar]$3d^9 4s$ 	&	$^1D_{2}$  	&	83500	\\
			&	30	&	9	&	Zn$^{3+}$	&	[Ar]$3d^9$    	&	$^2D_{5/2}$	&	0	&	    	&	31	&	10	&	Ga$^{3+}$	&	[Ar]$3d^{10}$ 	&	$^1S_{0}$  	&	0	\\
			&	   	&	  	&	         	&	[Ar]$3d^8 4s$ 	&	$^2F_{7/2}$	&	135951	&	    	&	   	&	  	&	         	&	[Ar]$3d^9 4s$ 	&	$^1D_{2}$  	&	156025	\\ \hline\hline
\end{tabular}}
\label{tab:3d-4s}
\end{table*}

\begin{table*}[t]
\scriptsize
\caption{The energy level of some ions in the Rb-like ($N=37$), Sr-like ($N=38$), and the fifth row transition metals from Y- to Pd-like ($N=39-46$) isoelectronic sequence HCIs, where $N$ is the number of the electron, $Z$ is the atomic number, and $x$ is the number of valence electron. [Kr]$\equiv 1s^2 2s^2 2p^6 3s^2 3p^6 3d^{10} 4s^2 4p^6$. The data are from the NIST database \cite{nist} and in cm$^{-1}$.}	
{\setlength{\tabcolsep}{5pt}
\begin{tabular}{lll lll l | lll lll l}\hline\hline
$N$	&	$Z$ 	&	$x $	&	Ion      	&	Configuration	&	State      	&	Energy	&	 $N$	&	$Z$ 	&	$x $	&	Ion      	&	Configuration	&	State      	&	Energy	\\ \hline
\multicolumn{7}{c}{Rb-like isoelectronic sequences} 													&	\multicolumn{7}{c}{Sr-like isoelectronic sequences}													\\
			37	&	38	&	1	&	Sr$^+$   	&	[Kr]$5s$     	&	$^2S_{1/2}$	&	0	&	38	&	39	&	2	&	Y$^{+}$  	&	[Kr]$5s^2$   	&	$^1S_{0}$  	&	0	\\
			&	  	&	  	&	         	&	[Kr]$4d$     	&	$^2D_{3/2}$	&	14555	&	  	&	  	&	  	&	         	&	[Kr]$4d^2$   	&	$^1D_{2}$  	&	14833	\\
			&	39	&	1	&	Y$^{2+}$ 	&	[Kr]$4d$     	&	$^2D_{3/2}$	&	0	&	  	&	40	&	2	&	Zr$^{2+}$	&	[Kr]$4d^2$   	&	$^3F_{2}$  	&	0	\\
			&	  	&	  	&	         	&	[Kr]$5s$     	&	$^2P_{1/2}$	&	7467	&	  	&	  	&	  	&	         	&	[Kr]$5s^2$   	&	$^1S_{0}$  	&	48506	\\
			&	40	&	1	&	Zr$^{3+}$	&	[Kr]$4d$     	&	$^2D_{3/2}$	&	0	&	  	&	41	&	2	&	Nb$^{3+}$	&	[Kr]$4d^2$   	&	$^3F_{2}$  	&	0	\\
			&	  	&	  	&	         	&	[Kr]$5s$     	&	$^2S_{1/2}$	&	38253	&	  	&	  	&	  	&	         	&	[Kr]$4d5s$   	&	$^3D_{1}$  	&	52064	\\ \hline
			\multicolumn{7}{c}{Y-like isoelectronic sequences}													&	\multicolumn{7}{c}{Zr-like isoelectronic sequences}													\\
			39	&	40	&	3	&	Zr$^{+}$ 	&	[Kr]$4d^2 5s$	&	$^4F_{3/2}$	&	0	&	40	&	41	&	4	&	Nb$^{+}$ 	&	[Kr]$4d^4$   	&	$^5D_{0}$  	&	0	\\
			&	  	&	  	&	         	&	[Kr]$4d 5s^2$	&	$^2D_{3/2}$	&	14298	&	  	&	  	&	  	&	         	&	[Kr]$4d^3 5s$	&	$^5F_{1}$  	&	2356	\\
			&	41	&	3	&	Nb$^{2+}$	&	[Kr]$4d^3$   	&	$^4F_{3/2}$	&	0	&	  	&	42	&	4	&	Mo$^{2+}$	&	[Kr]$4d^4$   	&	$^5D_{0}$  	&	0	\\
			&	  	&	  	&	         	&	[Kr]$4d^2 5s$	&	$^4F_{3/2}$	&	25220	&	  	&	  	&	  	&	         	&	[Kr]$4d^3 5s$	&	$^5F_{1}$  	&	32419	\\
			&	42	&	3	&	Mo$^{3+}$	&	[Kr]$4d^3$   	&	$^4F_{3/2}$	&	0	&	  	&	43	&	4	&	Tc$^{3+}$	&	[Kr]$4d^4$   	&	$^5D_{0}$  	&	0	\\
			&	  	&	  	&	         	&	[Kr]$4d^2 5s$	&	$^4F_{3/2}$	&	60896	&	  	&	  	&	  	&	         	&	[Kr]$4d^3 5s$	&	$^5F_{1}$  	&	71139	\\ \hline
			\multicolumn{7}{c}{Nb-like isoelectronic sequences}													&	\multicolumn{7}{c}{Mo-like isoelectronic s  sequences}													\\
			41	&	42	&	5	&	Mo$^{+}$ 	&	[Kr]$4d^5$   	&	$^6D_{5/2}$	&	0	&	42	&	43	&	6	&	Tc$^{+}$ 	&	[Kr]$4d^5 5s$	&	$^7S_{3}$  	&	0	\\
			&	  	&	  	&	         	&	[Kr]$4d^4 5s$	&	$^6D_{1/2}$	&	11783	&	  	&	  	&	  	&	         	&	[Kr]$4d^6$   	&	$^5D_{4}$  	&	3461	\\
			&	43	&	5	&	Tc$^{2+}$	&	[Kr]$4d^5$   	&	$^6D_{5/2}$	&	0	&	  	&	44	&	6	&	Ru$^{2+}$	&	[Kr]$4d^6$   	&	$^5D_{4}$  	&	0	\\
			&	  	&	  	&	         	&	[Kr]$4d^4 5s$	&	$^6D_{1/2}$	&	44254	&	  	&	  	&	  	&	         	&	[Kr]$4d^5 5s$	&	$^5S_{2}$  	&	41112	\\ \hline
			\multicolumn{7}{c}{Tc-like isoelectronic sequences} 													&	\multicolumn{7}{c}{Ru-like isoelectronic s  sequences} 													\\
			43	&	44	&	7	&	Ru$^{+}$ 	&	[Kr]$4d^7$   	&	$^4F_{9/2}$	&	0	&	44	&	45	&	8	&	Rh$^{+}$ 	&	[Kr]$4d^8$   	&	$^3F_{4}$  	&	0	\\
			&	  	&	  	&	         	&	[Kr]$4d^6 5s$	&	$^4D_{7/2}$	&	19379	&	  	&	  	&	  	&	         	&	[Kr]$4d^7 5s$	&	$^3F_{4}$  	&	25376	\\
			&	45	&	7	&	Rh$^{2+}$	&	[Kr]$4d^7$   	&	$^4F_{9/2}$	&	0	&	  	&	46	&	8	&	Pd$^{2+}$	&	[Kr]$4d^8$   	&	$^3F_{4}$  	&	0	\\
			&	  	&	  	&	         	&	[Kr]$4d^6 5s$	&	$^4D_{7/2}$	&	54632	&	  	&	  	&	  	&	         	&	[Kr]$4d^7 5s$	&	$^3F_{4}$  	&	62561	\\ \hline
			\multicolumn{7}{c}{Rh-like isoelectronic sequences}													&	\multicolumn{7}{c}{Pd-like isoelectronic s  sequences} 													\\
			45	&	46	&	9	&	Pd$^{+}$ 	&	[Kr]$4d^9$   	&	$^2D_{5/2}$	&	0	&	46	&	47	&	10	&	Ag$^{+}$ 	&	Kr]$4d^{10}$ 	&	$^1S_{0}$	&	0	\\
			&	  	&	  	&	         	&	[Kr]$4d^8 5s$	&	$^2F_{7/2}$	&	32281	&	  	&	  	&	  	&	         	&	Kr]$4d^9 5s$ 	&	$^3D_{3}$  	&	39168	\\
			&	47	&	9	&	Ag$^{2+}$	&	[Kr]$4d^9$   	&	$^2D_{5/2}$	&	0	&	  	&	48	&	10	&	Cd$^{2+}$	&	Kr]$4d^{10}$ 	&	$^1S_{0}$	&	0	\\
			&	  	&	  	&	         	&	[Kr]$4d^8 5s$	&	$^2F_{9/2}$	&	63246	&	  	&	  	&	  	&	         	&	Kr]$4d^9 5s$ &	$^3D_{3}$  	&	80454	\\ \hline\hline
\end{tabular}}
\label{tab:4d-5s}
\end{table*}

\begin{table*}[t]
\scriptsize
\caption{The energy levels of some ions in the isoelectronic (Isoelec.) sequences  of several sixth row transition metals, where $N$ is the number of the electron, $Z$ is the atomic number, and $x$ is the number of valence electron. [Xe]$\equiv 1s^2 2s^2 2p^6 3s^2 3p^6 3d^{10} 4s^2 4p^6 4d^{10} 5s^2 5p^6$. The data is taken from the NIST database unless cited others. Units for all the values are in cm$^{-1}$.}	
	{\setlength{\tabcolsep}{14pt}
		\begin{tabular}{cl ll ll ll}\hline\hline
Isoelec. Sequences	&	 $N$	&	$Z$ 	&	$x $	&	Ion      	&	Configuration	&	State      	&	Energy	\\ \hline
Tm-like	&	69	&	70	&	1	&	Yb$^{+}$ 	&	[Xe]$4f^{14}6s$      	&	$^2S_{1/2}$	&	0	\\
&	   	&	  	&	  	&	         	&	[Xe]$4f^{14}5d $     	&	$^2D_{3/2}$	&	22961	\\
Tm-like	&	69	&	71	&	1	&	Lu$^{2+}$	&	[Xe]$4f^{14}6s$      	&	$^2S_{1/2}$	&	0	\\
&	   	&	  	&	  	&	         	&	[Xe]$4f^{14}5d $     	&	$^2D_{3/2}$	&	5708	\\
Tm-like	&	69	&	72	&	1	&	Hf$^{3+}$	&	[Xe]$4f^{14}5d$      	&	$^2D_{3/2}$	&	0	\\
&	   	&	  	&	  	&	         	&	[Xe]$4f^{14}6s $     	&	$^2S_{1/2}$	&	18380 \cite{Allehabi2022-2}	\\
Yb-like	&	70	&	71	&	2	&	Lu$^{+}$ 	&	[Xe]$4f^{14} 6s^2$   	&	$^1S_{0}$  	&	0	\\
&	   	&	  	&	  	&	         	&	[Xe]$4f^{14} 5d^2$   	&	$^3P_{0}$  	&	35652	\\
Yb-like	&	70	&	72	&	2	&	Hf$^{2+}$	&	[Xe]$4f^{14} 5d^2$   	&	$^3F_{2}$  	&	0	\\
&	   	&	  	&	  	&	         	&	[Xe]$4f^{14} 5d6s$   	&	 $^3D_{1}$ 	&	2652 \cite{Berengut2011-2}	\\
Lu-like	&	71	&	72	&	3	&	Hf$^{+}$ 	&	[Xe]$4f^{14} 6s^2 5d$	&	$^2D_{3/2}$	&	0	\\
&	   	&	  	&	  	&	         	&	[Xe]$4f^{14} 6s 5d^2$	&	$^2F_{5/2}$	&	12070	\\
Lu-like	&	71	&	74	&	3	&	W$^{3+}$ 	&	[Xe]$4f^{14} 5d^3$   	&	$^4F_{3/2}$	&	0	\\
&	   	&	  	&	  	&	         	&	[Xe]$4f^{14} 5d^2 6s$	&	$^4F_{3/2}$	&	32692	\\
Ta-like	&	73	&	74	&	5	&	W$^{+}$  	&	[Xe]$4f^{14} 5d^4 6s$	&	$^6D_{1/2}$	&	0	\\
&	   	&	  	&	  	&	         	&	[Xe]$4f^{14} 5d^5$   	&	$^6S_{5/2}$	&	7420	\\
Ta-like	&	73	&	75	&	5	&	Re$^{2+}$	&	[Xe]$4f^{14} 5d^5$   	&	$^6S_{5/2}$	&	0	\\
&	   	&	  	&	  	&	         	&	[Xe]$4f^{14} 5d^4 6s$	&	$^6D_{1/2}$	&	143961	\\
W-like	&	74	&	75	&	6	&	Re$^{+}$ 	&	[Xe]$4f^{14} 5d^5 6s$	&	$^7S_{3}$  	&	0	\\
&		&		&		&		&	[Xe]$4f^{14} 5d^4 6s^2$	&	$^5D_{0}$  	&	13777	\\
W-like 	&	74	&	76	&	6	&	Os$^{2+}$	&	[Xe]$4f^{14} 5d^6$	&	$^5D_{4}$  	&	0	\\
&	   	&	  	&	  	&	         	&	[Xe]$4f^{14} 5d^56s   $	&	$^7S_{3}$  	&	4578	\\
W-like	&	74	&	77	&	6	&	Ir$^{3+}$	&	[Xe]$4f^{14} 5d^6$   	&	$^5D_{4}$  	&	0	\\
&	   	&	  	&	  	&	         	&	[Xe]$4f^{14} 5d^6$	&	$^3H_{4}$  	&	14546	\\
Ir-like	&	77	&	78	&	9	&	Pt$^{+}$ 	&	[Xe]$4f^{14} 5d^9$   	&	$^2D_{5/2}$	&	0	\\
&	   	&	  	&	  	&	         	&	[Xe]$4f^{14} 5d^8 6s$	&	$^4F_{9/2}$	&	4787	\\
Ir-like	&	77	&	79	&	9	&	Au$^{2+}$	&	[Xe]$4f^{14} 5d^9$   	&	$^2D_{5/2}$	&	0	\\
&	   	&	  	&	  	&	         	&	[Xe]$4f^{14} 5d^8 6s$	&	$^4F_{9/2}$	&	50818 \cite{Zainab2019}   	\\ \hline\hline			
\end{tabular}}
\label{tab:5d-6s}
\end{table*}

\subsubsection{The $nd-(n+1)s$ level crossings} \label{lcross}

Table \ref{tab:3d-4s} summarizes the $3d-4s$ energy level-crossing occurring in the K-like and Ca-like isoelectronic sequences and the isoelectronic sequences of the elements belonging to the $3d$ metals. The ground state configurations of the singly charged ions belonging to these elements are generally $3d^{x-1}4s$, while the ground state configuration of their double charged ions happens to be $3d^{x}$ with $x$ denoting the number of the valence electrons that is one ($x=1$) for the K-like isoelectronic sequence, two ($x=2$) for the Ca-like isoelectronic sequences, and so on till $x=10$ for the Ni-like isoelectronic sequences. So one would anticipate that the $3d-4s$ energy level crossing can occur in these isoelectronic sequences when the charges of the ions increase further. However, there have been several exceptions seen where no $3d-4s$ level-crossings have occurred in these isoelectronic sequences, such as HCIs belonging to the Ti-like and V-like isoelectronic sequences which have the number of the valence electrons close to half-filling and to the Fe-, Co-, and Ni-like isoelectronic sequences that are close to full-filling. In these cases, the ground state configuration is preferred to be $3d^{x}$. Table \ref{tab:4d-5s} summarizes the $4d-5s$ energy level-crossings that occur in a series of isoelectronic sequences of elements from the fifth row of the periodic table. The $4d-5s$ energy level-crossings are observed in the Rb-, Sr-, and Y-like isoelectronic sequences with $x=1-3$ and in the Mo-like isoelectronic sequence with $x=6$. However, no such energy level crossings are seen in the isoelectronic sequences with the number of valence electrons close to half- or fully-filled. i.e., for $x=4,5,7,8,9,10$. For the elements belonging to the sixth row of the periodic table, the $5d-6s$ energy level crossings are observed in some isoelectronic sequences, shown in Table \ref{tab:5d-6s}, such as in the Tm-, Yb-, Lu-, Ta-, W-, and Ir-like isoelectronic sequences.

As shown in Tables \ref{tab:3d-4s}, \ref{tab:4d-5s}, and \ref{tab:5d-6s}, the $nd-(n+1)s$ energy level-crossings happen often in the doubly-charged ions. In these doubly-charged ions, transition wavelengths between their ground and first excited states in these ions are within the optical range. Such transitions are accompanied by changes in the electronic angular momenta from the $s$ to $d$ orbitals. Thus, considering these HCIs as atomic clocks would offer substantial relativistic sensitivity coefficients to probe $\alpha$ variation. For example, The Ac$^{2+}$, Zr$^{2+}$, Hf$^{2+}$, and Hg$^{2+}$ has been suggested as suitable candidate for laboratory searches of space-time variation of $\alpha$ \cite{Berengut2011-2}. We find that many double charged HCI ions have similar transitions, such as the $3d^2-3d4s$ transition in Ti$^{2+}$, the $3d^3-3d^24s$ transition in V$^{2+}$, the $4d^3-4d^25s$ transition in Nb$^{2+}$, the $5d^5-5d^46s$ transition in Re$^{2+}$, the $5d^6-5d^56s$ transition in Os$^{2+}$, and the $5d^9-5d^86s$ transition in Au$^{2+}$, which could be the promising HCIs for the study of possible temporal variation of $\alpha$. The singly and triply charged ions also have some new interesting optical transitions, for example, Allehabi, et al. have demonstrated that the metastable excited state in Hf$^{+}$ and Hf$^{3+}$ may be good clock states having sufficiently long-lived upper states, insensitive to the external perturbations, and additional E1 cooling line available \cite{Allehabi2022-2}. 

\begin{table*}[t]
	\scriptsize
	\caption{The energy level of HCIs in the isoelectronic sequences containing the different number of $5s$ and $4f$ valence electrons $x$, where $N$ is the number of the electron, $Z$ is the atomic number. [Kr]$\equiv 1s^2 2s^2 2p^6 3s^2 3p^6 3d^{10} 4s^2 4p^6$. Units for values are in cm$^{-1}$. }	
	{\setlength{\tabcolsep}{12pt}
		\begin{tabular}{cl ll ll ll }\hline\hline
			Isoelec.	&	 $N$	&	$Z$ 	&	$x $	&	Ion      	&	Configuration	&	State      	&	Energy	\\ \hline
			Ag-like	&	47	&	60	&	1	&	Nd$^{13+}$	&	[Kr]$ 4d^{10} 5s $           	&	$^2S_{1/2}$    	&	0	\\
			&	  	&	  	&	  	&	          	&	[Kr]$ 4d^{10} 4f^1 $         	&	$^2F^o_{5/2}$  	&	55706 \cite{Safronova2014Ag}   	\\
			Ag-like	&	47	&	61	&	1	&	Pm$^{14+}$	&	[Kr]$ 4d^{10} 5s $           	&	$^2S_{1/2}$    	&	0	\\
			&	  	&	  	&	  	&	          	&	[Kr]$ 4d^{10} 4f^1 $         	&	$^2F^o_{5/2}$  	&	8902 \cite{Berengut2012-4}       	\\
			Ag-like	&	47	&	62	&	1	&	Sm$^{15+}$	&	[Kr]$ 4d^{10} 4f^1 $         	&	$^2F^o_{5/2}$  	&	0	\\
			&	  	&	  	&	  	&	          	&	[Kr]$ 4d^{10} 5s $           	&	$^2S_{1/2}$    	&	60517 \cite{Berengut2012-4}      	\\
			Cd-like	&	48	&	61	&	2	&	Pm$^{13+}$	&	[Kr]$ 4d^{10} 5s^2 $         	&	$^1S_{0}$      	&	0	\\
			&	  	&	  	&	  	&	          	&	[Kr]$ 4d^{10} 5s 4f$         	&	$^3F^o_{2}$    	&	86136 \cite{Berengut2012-4}      	\\
			Cd-like	&	48	&	62	&	2	&	Sm$^{14+}$	&	[Kr]$ 4d^{10} 4f^2 $         	&	$^3H_{4}$      	&	0	\\
			&	  	&	  	&	  	&	          	&	[Kr]$ 4d^{10} 5s 4f$         	&	$^3F^o_{2}$    	&	2172 \cite{Safronova2014Cd}    	\\
			Cd-like	&	48	&	63	&	2	&	Eu$^{15+}$	&	[Kr]$ 4d^{10} 4f^2 $         	&	$^3H_{4}$      	&	0	\\
			&	  	&	  	&	  	&	          	&	[Kr]$ 4d^{10} 5s 4f$         	&	$^3F^o_{2}$    	&	48780 \cite{Berengut2012-4}      	\\
			In-like	&	49	&	62	&	3	&	Sm$^{13+}$	&	[Kr]$ 4d^{10} 5s^2  4f$      	&	$^2F^o_{5/2}$  	&	0	\\
			&	  	&	  	&	  	&	          	&	[Kr]$ 4d^{10} 5s  4f^2$      	&	$^4H_{7/2}$    	&	20254 \cite{Safronova2014Ag}   	\\
			In-like	&	49	&	63	&	3	&	Eu$^{14+}$	&	[Kr]$ 4d^{10} 4f^2 5s $      	&	$J=3.5$        	&	0	\\
			&	  	&	  	&	  	&	          	&	[Kr]$ 4d^{10} 4f 5s^2 $      	&	$J=3.5$ (odd)  	&	24854 \cite{Berengut2012-4}      	\\
			In-like	&	49	&	64	&	3	&	Gd$^{15+}$	&	[Kr]$ 4d^{10} 4f^{3}  $      	&	$J=4.5$ (odd)  	&	0	\\
			&	  	&	  	&	  	&	          	&	[Kr]$ 4d^{10} 4f^{2}5s$      	&	$J=3.5$        	&	30172  \cite{Berengut2012-4}     	\\
			I-like	&	53	&	67	&	7	&	Ho$^{14+}$	&	 [Kr]$ 4d^{10} 5s   4f^6$     	&	 $^8F_{1/2}$  	&	0	\\
			&	  	&	  	&	  	&	          	&	 [Kr]$ 4d^{10} 5s^2 4f^5$     	&	 $^6H^o_{5/2}$	&	23800\cite{Dzuba2015}       	\\
			Xe-like	&	54	&	68	&	8	&	Er$^{14+}$	&	 [Kr]$ 4d^{10} 5s 4f^7  $     	&	$^9S^o_{4}$   	&	0	\\
			&	  	&	  	&	  	&	          	&	 [Kr]$ 4d^{10} 5s^2 4f^6$     	&	$^7F_{0}$     	&	18555 \cite{Dzuba2015}      	\\
			Ce-like	&	58	&	74	&	12	&	W$^{16+}$ 	&	 [Kr]$ 4d^{10} 5s 4f^{11} $      	&	$^5I^o_{8}$    	&	0	\\
			&		&		&		&		&	 [Kr]$ 4d^{10} 5s^2 4f^{10} $      	&	$^5I_{8}$    	&	19000 \cite{nist}	\\
			Pr-like	&	59	&	77	&	13	&	Ir$^{18+}$	&	Kr]$ 4d^{10} 4f^{13}     $       	&	$^2F^o_{7/2}$ 	&	0	\\
			&	  	&	   	&	 	&	         	&	Kr]$ 4d^{10} 5s 4f^{12} $        	&	$^4F_{9/2}$  	&	70096 \cite{Safronova2015}	\\
			Nd-like	&	60	&	74	&	14	&	Os$^{16+}$	&	[Kr]$ 4d^{10} 5s^2 4f^{12} $       	&	$^3H_{6}$     	&	0 	\\
			&		&		&		&		&	[Kr]$ 4d^{10} 5s 4f^{13} $       	&	$^3F^o_{4}$     	&	37460  \cite{Windberger2015}	\\
			Nd-like	&	60	&	77	&	14	&	Ir$^{17+}$	&	[Kr]$ 4d^{10} 5s 4f^{13} $       	&	$^3F^o_{4}$   	&	0 	\\
			&	  	&	   	&	  	&	          	&	[Kr]$ 4d^{10} 4f^{14}      $     	&	$^1S_{0}$     	&	13530  \cite{Windberger2015}	\\
			Nd-like	&	60	&	78	&	14	&	Pt$^{18+}$	&	[Kr]$ 4d^{10} 4f^{14}      $     	&	$^1S_{0}$     	&	0 	\\
			&	  	&	   	&	  	&	          	&	[Kr]$ 4d^{10} 5s 4f^{13}   $     	&	$^3F^o_{4}$   	&	48809  \cite{Windberger2015}	\\
			Pm-like	&	61	&	74	&	15	&	W$^{13+}$	&	[Kr]$ 4d^{10} 5s^2 4f^{13} $     	&	$^2F^o_{7/2}$ 	&	0	\\
			&	  	&	   	&	  	&	          	&	[Kr]$ 4d^{10} 5s^2 4f^{13} $     	&	$^2F^o_{5/2}$ 	&	26000 \cite{nist,Liu2022}	\\
			Sm-like	&	61	&	77	&	15	&	Ir$^{16+}$	&	[Kr]$ 4d^{10} 5s^2 4f^{13} $     	&	$^2F^o_{7/2}$ 	&	0	\\
			&	  	&	   	&	  	&	          	&	[Kr]$ 4d^{10} 5s^2 4f^{13} $     	&	$^2F^o_{5/2}$ 	&	25023 \cite{Nandy2016}	\\
			Sm-like	&	61	&	78	&	15	&	Pt$^{17+}$	&	[Kr]$ 4d^{10} 5s  4f^{14} $     	&	$^2S_{1/2}$ 	&	0	\\
			&	  	&	   	&	  	&	          	&	[Kr]$ 4d^{10} 5s^2 4f^{13} $     	&	$^2F^o_{7/2}$ 	&	24860 \cite{Nandy2016}	\\ \hline\hline
	\end{tabular}} \label{tab:4f-5s}
\end{table*}

\begin{table*}[t]
	\scriptsize
	\caption{The energy level of HCIs in the isoelectronic sequences containing the different number of $5p$ and $4f$ valence electrons $x$, where $N$ is the number of the electron, $Z$ is the atomic number. [Kr]$\equiv 1s^2 2s^2 2p^6 3s^2 3p^6 3d^{10} 4s^2 4p^6$ and [Cd]$\equiv 1s^2 2s^2 2p^6 3s^2 3p^6 3d^{10} 4s^2 4p^6 4d^{10} 5s^2$. Units of all values are in cm$^{-1}$.}	
	{\setlength{\tabcolsep}{13pt}
		\begin{tabular}{cl ll ll ll}\hline\hline
			Isoelec.&	 $N$	&	$Z$ 	&	$x $	&	Ion     &	Configuration	&	State  &Energy	\\ \hline
			In-like	&	49	&	58	&	1	&	Ce$^{11+}$	&	[Kr]$ 5p$ 	&	$^2P^{\circ}_{1/2}$	&	0	\\
			&		&		&		&		&	[Kr]$ 5p$ 	&	$^2P^{\circ}_{3/2}$	&	33427 \cite{Safronova2014Ag}	\\
			&		&		&		&		&	[Kr]$ 4f$ 	&	$^2F^{\circ}_{5/2}$	&	54947 \cite{Safronova2014Ag}	\\
			In-like	&	49	&	59	&	1	&	Pr$^{10+}$	&	[Kr]$ 5p$ 	&	$^2P^{\circ}_{1/2}$	&	0	\\
			&		&		&		&		&	[Kr]$ 4f$ 	&	$^2F^{\circ}_{5/2}$	&	3958 \cite{Safronova2014Ag}	\\
			&		&		&		&		&	[Kr]$ 5p$ 	&	$^2P^{\circ}_{3/2}$	&	39084\cite{Safronova2014Ag}	\\
			In-like	&	49	&	60	&	1	&	Nd$^{11+}$	&	[Kr]$ 4f$ 	&	$^2F^{\circ}_{5/2}$	&	0	\\
			&		&		&		&		&	[Kr]$ 4f$ 	&	$^2F^{\circ}_{7/2}$	&	4566 \cite{Safronova2014Ag}	\\
			&		&		&		&		&	[Kr]$ 5p$ 	&	$^2P^{\circ}_{1/2}$	&	53491\cite{Safronova2014Ag}	\\
			Sn-like	&	50	&	59	&	2	&	Pr$^{9+}$	&	[Kr]$ 5p^2$ 	&	$^3P_{o}$ 	&	0	\\
			&		&		&		&		&	[Kr]$ 5p 4f$ 	&	$^3G_{3}$ 	&	22101 \cite{Bekker2019}	\\
			&		&		&		&		&	[Kr]$ 5p^2$ 	&	$^3P_{1}$ 	&	28561  \cite{Bekker2019}	\\
			Sn-like	&	50	&	60	&	2	&	Nd$^{10+}$	&	[Kr]$ 4f^2$ 	&	$^3H_{4}$ 	&	0	\\
			&		&		&		&		&	[Kr]$ 5p 4f$ 	&	$^3G_{3}$ 	&	2605  \cite{Safronova2014Cd}	\\
			&		&		&		&		&	[Kr]$ 4f^2$ 	&	$^3H_{5}$ 	&	3432  \cite{Safronova2014Cd}	\\
			Sb-like	&	51	&	59	&	3	&	Pr$^{8+}$	&	[Kr]$ 5p^3$ 	&	$^4S^{\circ}_{3/2}$ 	&	0	\\
			&		&		&		&		&	[Kr]$ 5p^2 4f$ 	&	$^4G^{\circ}_{5/2}$ 	&	18386 \cite{Huo2017}	\\
			&		&	  	&	  	&		&	[Kr]$ 5p^3$ 	&	$^2D_{3/2}$ 	&	24348 \cite{Huo2017}	\\
			Sb-like	&	51	&	60	&	3	&	Nd$^{9+}$	&	[Kr]$ 5p^2 4f$ 	&	$J=5/2$ 	&	0	\\
			&		&		&		&		&	[Kr]$ 5p^2 4f$ 	&	$J=7/2$	&	6524  \cite{Yu2020}	\\
			&	  	&	  	&	  	&	          	&	[Kr]$ 4f^2 5p $ 	&	$J=9/2$ 	&	20594  \cite{Yu2020}	\\
			Sb-like	&	51	&	61	&	3	&	Pm$^{10+}$	&	[Kr]$ 4f^2 5p$ 	&	$J=9/2$ 	&	0	\\
			&		&		&		&		&	[Kr]$ 4f^3 $ 	&	$J=9/2$ 	&	5296 \cite{Huo2017}	\\
			Sb-like	&	51	&	62	&	3	&	Sm$^{11+}$	&	[Kr]$ 4f^3$ 	&	$^4I^{\circ}_{9/2}$ 	&	0	\\
			&		&		&		&		&	[Kr]$ 4f^3 $ 	&	$^4I^{\circ}_{11/2}$ 	&	3063 \cite{Huo2017}	\\
			Sm-like	&	62	&	74	&	14	&	W$^{12+}$ 	&	[Cd]$ 4f^{14}   $ 	&	$^1S_{0}$	&	0	\\
			&	  	&	  	&	  	&	          	&	[Cd]$ 4f^{13} 5p$ 	&	$^3D_{3}$	&	4016 \cite{Lu2022}	\\
			Eu-like	&	63	&	74	&	15	&	W$^{11+}$ 	&	[Cd]$ 4f^{13} 5p^2 $	&	$^4F^o_{7/2}$ 	&	0	\\
			&	  	&	  	&	  	&	          	&	[Cd]$ 4f^{14} 5p$   	&	$^2P^{\circ}_{1/2 }$	&	11000 \cite{nist,Singh2020,Fu2022}	\\
			Gd-like	&	64	&	74	&	16	&	W$^{10+}$	&	[Cd]$ 4f^{14}5p^2 $	&	$^3P_{2}$ 	&	0	\\
			&		&		&		&		&	[Cd]$ 4f^{13}5p^3$	&	$^3G_{5}$ 	&	44614 \cite{Lu2021}	\\
			Tb-like	&	65	&	74	&	17	&	W$^{9+}$	&	[Cd]$ 4f^{14}5p^3 $	&	$^2P^{\circ}_{3/2}$ 	&	0	\\
			&		&		&		&		&	[Cd]$ 4f^{13}5p^4$	&	$^3G^{\circ}_{5}$ 	&	44614 \cite{Priti2020}	\\
			Dy-like	&	66	&	74	&	18	&	W$^{8+}$	&	[Cd]$ 4f^{14}5p^4 $	&	$^3P_{2}$ 	&	0	\\
			&		&		&		&		&	[Cd]$ 4f^{13}5p^5$	&	$^3F_{4}$ 	&	14000 \cite{nist,Priti2020} 	\\ \hline\hline
	\end{tabular}}
	\label{tab:4f-5p}
\end{table*}

\subsubsection{The $4f-5s$ and $4f-5p$ level crossings}
The $4f-5s$ orbital energy crossing frequently occurs in the isoelectronic sequences of the electronic configuration containing open-shelled $5s$ 
and $4f$ orbital. Table \ref{tab:4f-5s} lists some HCIs in such kinds of isoelectronic sequences. This type of HCIs usually has the large number of 
charge, larger than 10 in most cases. The Ag-, Cd-, In-, Sn-like HCIs containing one, two, three, and four valence electrons have been investigated
by Safronova, et al \cite{Safronova2014,Safronova2014Ag, Safronova2014Cd}. When valence electrons are larger than 4, the electronic configuration becomes far more complicated. Therefore there is little data available for such kind of HCIs. Dzuba has investigated the energy levels 
of the I-like Ho$^{14+}$ and Xe-like Er$^{15+}$ with $x$= 6 and 7, which are proposed as candidates for an extremely accurate and stable optical 
atomic clock which is found to have large values of $q$ and $K$ suitable for the study of time variation of $\alpha$. When $x\geq12$, there observed 
some HCIs having the $4f-5s$ crossing in the hole states of the $5s$ and $4f$ shells, as listed in Table \ref{tab:4f-5s}. For example, the 
Ir$^{17+}$ ion has been proposed for the search of variation of $\alpha$, and its hole state of $5s4f^{14}$ has extremely high values of $q$ and $K$,
indicating the hole states can dramatically enhance the sensitivity to $\alpha$ variation \cite{Berengut2011}. Nandy and Sahoo have also 
investigated the W$^{13+}$, Ir$^{16+}$ and Pt$^{17+}$ HCIs with ground state configurations $4f^{13}$ as promising optical clock candidates to probe
variation of $\alpha$ \cite{Nandy2016}.

The $4f-5p$ energy level-crossing occurs in the isoelectronic sequences containing the open-shelled $5p$ and $4f$ configurations, and this type of HCIs usually have mediated number of charge, around ten or lower. Table \ref{tab:4f-5p} lists some of these kinds of HCIs. Safronova et al. have 
proposed  In-like Pr$^{10+}$ and Sn-like Pr$^{9+}$ and Nd$^{10+}$ as the potential candidates for atomic clocks \cite{Safronova2014Ag,
	Safronova2014Cd,Safronova2014}. Berengnut et al. have also proposed Sn-like Pr$^{9+}$ as a promising candidate for developing an extremely 
accurate atomic clock for the measurement of sensitivity to variation of $\alpha$. Yu et al. have calculated the atomic energy levels of Sb-like 
Nd$^{9+}$ and suggested Nd$^{9+}$ as a potential candidate for making an atomic clock \cite{Yu2020}. The In-, Sn-, and Sb-like HCIs with 
mediated number of charge contain one, two, 
and three valence electrons distributing on the $5p$ and $4f$ shells. However, accurate calculation of such systems also needs to take into two inner occupied $5s$ electrons, which makes 
multi-valent systems containing 3-5 valence electrons distributing on 11 orbitals of $5s$, $5p$, and $4f$. Their complex electronic configurations 
make studying spectroscopic properties from theoretical and experimental viewpoints challenging. Due to the scarcity of data on energy 
level, isoelectronic sequences with $x$ larger than 3 are not in Table \ref{tab:4f-5p}. When $x\geq$ 14, there observed the hole state of $4f$ shell 
in some HCIs, for example W$^{(8-13)+}$ ions, as listed in \ref{tab:4f-5p}. Their $4f$- or $5p-$ excitation causes a lot of transitions with energies in the 
optical range. There has been greatly renewed interest in the spectral emission of tungsten from high-temperature plasma 
\cite{Priti2020,Safronova2011,Lu2021,Lu2021-2,Kozol2021,Lu2022,Fu2022}.

Tables \ref{tab:3d-4s}, \ref{tab:4d-5s}, \ref{tab:5d-6s}, \ref{tab:4f-5s}, and \ref{tab:4f-5p} compile a series of HCIs that underline the 
orbital energy-crossings in their electronic configurations. Many of these HCIs show optical transitions between the ground and low-lying excited 
states. Although the exact values of $q$ and $K$ for these transitions are mostly absent in the literature, their values can be conjectured to be 
very large from the characterstics of the $s-f$ and $p-f$ inter-configuration transitions. Such features in atomic clocks are useful for  
probing $\alpha$ variation. It is, therefore, necessary to estimate energies and other spectroscopic properties of these relevant for 
estimating systematic errors if they are undertaken for atomic clocks.

\section{Relativistic Many-body Methods}

There is a great need to determine the energies and radiative properties of the HCIs to determine their suitability for atomic 
clocks. Again, their high-accuracy estimations are also essential for gauging the atomic clocks' systematic effects.  
The required atomic data include energy level structures, atomic polarizabilities, hyperfine structure constants, Land\'{e} $g_J$ factors, quadrupole
moments, etc., which are unknown for many HCIs. For theoretical evaluation of these quantities, it is necessary to employ methods that can reliably 
estimate them. Though relativistic contributions from the Breit and quantum electrodynamics (QED) effects would be significant in the HCIs, Coulomb interactions still play decisive roles in accurately determining properties of the interested HCIs for finding their aptness to make atomic clocks. Thus, we adopt relativistic many-body methods that can estimate atomic properties with reasonable accuracy in HCIs, and corrections from the higher-order 
relativistic effects are estimated approximately wherever required. Again, it is impossible to employ a single many-body method to all the HCIs 
undertaken here for investigation due to their multi-valence electronic configurations. We choose a method that can be applied aptly in HCIs with a 
particular electronic configuration class. In this view, we have considered methods like the configuration interaction (CI) method, general active 
space configuration interaction (GASCI) method and Fock-space relativistic coupled-cluster (FS-RCC) theory in the present work. In all the
methods, wave functions obtained using the Dirac-Hartree-Fock (DHF) method are considered the starting point. Correlation effects due to the 
residual Coulomb interactions are included through the many-body approaches.

\subsection{Approximations in the Hamiltonian}

To take into account the major relativistic effects and electron correlation effects in the HCIs, we first consider the Dirac-Coulomb (DC)
Hamiltonian in our calculations which in atomic units (a.u.) is given by
\begin{eqnarray}\label{eq:DHF}
	H^{DC} &=& \sum_i \left [c\mbox{\boldmath$\alpha$}_i\cdot \textbf{p}_i+(\beta_i-1)c^2+V_{nuc}(r_i)\right] +\sum_{i,j>i}\frac{1}{r_{ij}} \nonumber \\
	&=& \sum_i h_i + \sum_{i,j>i} g_{ij} ,
\end{eqnarray}
where the rest mass energies of the electrons are subtracted. In this expression $c$ is the speed of light, $V_{nuc}(r)$ is the nuclear 
potential and $\mbox{\boldmath$\alpha$}=(\alpha_x, \alpha_y, \alpha_z)$  and $\beta$ are the $4\times4$ Dirac matrices with the components
\begin{equation}
	\alpha_x= \bigg( \begin{array}{c c c} 0_2&\sigma_x\\\sigma_x&0_2  \end{array} \bigg), \alpha_y= 
	\bigg( \begin{array}{c c c} 0_2&\sigma_y\\\sigma_y&0_2  \end{array} \bigg), 
	\alpha_z= \bigg( \begin{array}{c c c} 0_2&\sigma_z\\\sigma_z&0_2  \end{array} \bigg) \ \ \text{and} \ \ \beta= \bigg( \begin{array}{c c c} I&0\\0&I  \end{array} \bigg),
\end{equation} 
for the identity matrix $I$ and the Pauli spin matrices 
\begin{equation}
	\sigma_x= \bigg( \begin{array}{c c}0&1\\1&0 \end{array} \bigg), 
	\sigma_y= \bigg( \begin{array}{c c} 0&-i\\i&0  \end{array} \bigg) \ \ \text{and} \ \ \sigma_z= \bigg( \begin{array}{c c} 1&0\\0&-1  \end{array} \bigg) .
\end{equation} 

In our calculations, we adopt either the Gaussian or Fermi nuclear charge distribution \cite{Visscher1997} to obtain the nuclear potentials of the heavier HCIs. The Gaussian nuclear charge distribution is defined by 
\begin{eqnarray}
	\rho(r) = \left(\frac{\eta}{\pi}\right)^{\frac{3}{2}} e^{-\eta r^2}
\end{eqnarray}
with $\eta=\frac{3}{2}R_{rms}^{-2}$, where $R_{rms}$ is the root mean square (rms) nuclear charge radius of the atomic nucleus. This leads to the 
expression for nuclear potential observed by an electron as 
\begin{eqnarray}
	V_{nuc}(r)=- \frac{Z}{r} \ erf\left ( \sqrt{\eta} r\right) .
\end{eqnarray}
In the Fermi nuclear charge distribution, it is given as
\begin{eqnarray}
	\rho(r) = \frac{\rho_0}{1+e^{(r-b)/a}} ,
\end{eqnarray}
where $\rho_0$ is the normalization constant, $b$ is the half-charge radius and $a=2.3/4\ln{3}$ is known as the skin thickness. We obtain
the $b$ value using the relation 
\begin{eqnarray}
	b&=& \sqrt{\frac {5}{3} R_{rms}^2 - \frac {7}{3} a^2 \pi^2}
\end{eqnarray}
and rms charge radius of a given nucleus with atomic mass $A$ is estimated in $fm$ by
\begin{eqnarray}
	R_{rms} =0.836 A^{1/3} + 0.570 .
\end{eqnarray}
The above expression gives the nuclear potential expression as
\begin{eqnarray}
	V_{nuc}(r) = -\frac{Z}{\mathcal{N}r} 
	\left\{\begin{array}{rl}
		\frac{1}{b}(\frac{3}{2}+\frac{a^2\pi^2}{2b^2}-\frac{r^2}{2b^2}+\frac{3a^2}{b^2}P_2^+\frac{6a^3}{b^2r}(S_3-P_3^+)) & \mbox{for $r_i \leq b$}\\
		\frac{1}{r_i}(1+\frac{a62\pi^2}{b^2}-\frac{3a^2r}{b^3}P_2^-+\frac{6a^3}{b^3}(S_3-P_3^-))                           & \mbox{for $r_i >b$} ,
	\end{array}\right.   
	\label{eq12}
\end{eqnarray}
where the factors are 
\begin{eqnarray}
	\mathcal{N} &=& 1+ \frac{a^2\pi^2}{b^2} + \frac{6a^3}{b^3}S_3 \ \ \text{with} \ \ S_k = \sum_{l=1}^{\infty} \frac{(-1)^{l-1}}{l^k}e^{-lb/a} \ \
	\text{and} \ \ P_k^{\pm} = \sum_{l=1}^{\infty} \frac{(-1)^{l-1}}{l^k}e^{\pm l(r-b)/a} . 
\end{eqnarray}

Whenever necessary, we add the potential ($V^B$) due to the Breit interaction to the DC Hamiltonian to take into account the contribution from this 
higher-order relativistic effect and is given by
\begin{eqnarray}\label{eq:DHB}
	V^B  &=& - \sum_{j>i}\frac{[\mbox{\boldmath$\alpha$}_i\cdot\mbox{\boldmath$\alpha$}_j+
		(\mbox{\boldmath$\alpha$}_i\cdot\mathbf{\hat{r}_{ij}})(\mbox{\boldmath$\alpha$}_j\cdot\mathbf{\hat{r}_{ij}})]}{2r_{ij}} ,
\end{eqnarray}
where $\mathbf{\hat{r}_{ij}}$ is the unit vector along the inter-electronic distance $\mathbf{r_{ij}}$.

Similarly, contributions from the QED effects are estimated approximately by considering the lower-order vacuum polarization (VP) interaction 
($V_{VP}$) and the self-energy (SE) interactions ($V_{SE}$) through the model potentials. We account for $V_{VP}$ through the Uehling  
and Wichmann-Kroll potentials ($V_{VP}=V^{Uehl} + V^{WK}$), given by
\begin{eqnarray}
	\label{eq:uehl}
	V^{Uehl}&=&- \frac{2}{3} \sum_i \frac{\alpha^2 }{r_i} \int_0^{\infty} dx \ x \ \rho_n(x)\int_1^{\infty}dt \sqrt{t^2-1} 
	\left(\frac{1}{t^3}+\frac{1}{2t^5}\right)  \left [ e^{-2ct|r_i-x|} - e^{-2ct(r_i+x)} \right ]\ \ 
\end{eqnarray}
and
\begin{eqnarray}
	V^{WK} = \sum_i \frac{0.368 Z^2}{9 \pi c^3 (1+(1.62 c r_i )^4) } \rho(r_i),
\end{eqnarray}
respectively. The SE contribution $V_{SE}$ is estimated by including two parts as 
\begin{eqnarray}
	V_{SE}^{ef}&=&  A_l \sum_i \frac{2 \pi Z \alpha^3 }{r_i} I_1^{ef}(r_i) - B_l \sum_i \frac{\alpha }{ r_i} I_2^{ef}(r_i) \ \ \
\end{eqnarray}
known as effective electric form factor part and
\begin{eqnarray}
	V_{SE}^{mg}&=& - \sum_k \frac{i\alpha^3}{4} \mbox{\boldmath$\gamma$} \cdot \mbox{\boldmath$\nabla$}_k \frac{1}{r_k} 
	\int_0^{\infty} dx \ x \ \rho_n(x) \int_1^{\infty} dt \frac{1}{t^3 \sqrt{t^2-1}} \nonumber \\
	&& \times \left [ e^{-2ct|r_k-x|} - e^{-2ct(r_k+x)} - 2ct \left (r_k+x-|r_k-x| \right ) \right ], \nonumber \\
\end{eqnarray}
known as the effective magnetic form factor part. In the above expressions, we use 
\begin{eqnarray}
	A_l= \begin{cases} 0.074+0.35Z \alpha \ \text{for} \ l=0,1 \\  0.056+0.05 Z \alpha + 0.195 Z^2 \alpha^2 \ \text{for} \ l=2  , \end{cases}
\end{eqnarray}
and
\begin{eqnarray}
	B_l = \begin{cases} 1.071-1.97x^2 -2.128 x^3+0.169 x^4  \ \text{for} \ l=0,1 \\
		0 \ \text{for} \ l \ge 2 .  \end{cases}   
\end{eqnarray}
The integrals are given by
\begin{eqnarray}
	I_1^{ef}(r) =  \int_0^{\infty} dx \ x \ \rho_n(x) [ (Z |r-x|+1) e^{-Z|r-x|}  - (Z(r+x)+1) e^{-2ct(r+x)}  ] 
\end{eqnarray}
and
\begin{eqnarray}
	I_2^{ef}(r) = \int_0^{\infty} dx \ x \ \rho_n(x)  \int^{\infty}_1 dt \frac{1}{\sqrt{t^2-1}} \bigg \{ \left( 1-\frac{1}{2t^2} \right ) 
	\left [ \ln(t^2-1)+4 \ln \left ( \frac{1}{Z \alpha} +\frac{1}{2} \right ) \right ]-\frac{3}{2}+\frac{1}{t^2} \big \} && \nonumber \\
	\times  \{  \frac{\alpha}{t} \left [ e^{-2ct|r-x|} - e^{-2ct(r+x)} \right ] +2 r_A e^{2 r_A ct } 
	\left [ E_1 (2ct (|r-x|+r_A)) - E_1 (2ct (r+x+r_A)) \right ] \bigg \}  &&
\end{eqnarray}
with the orbital quantum number $l$ of the system, $x=(Z-80)\alpha$, $r_A= 0.07 Z^2 \alpha^3$, and the exponential integral $E_1(r) = 
\int_r^{\infty} ds e^{-s}/s$.

The atomic Hamiltonian given by Eq. (\ref{eq:DHF}) can be expressed in the second-quantization formalism as
\begin{equation}
	H^{DC}= \sum_{pq}h^p_qa^{+}_pa_q+\frac{1}{4}\sum_{pqrs}V^{pr}_{qs}a^+_pa^+_ra_sa_q,
\end{equation}
where $h^p_q=\langle p|\hat{h}|q\rangle$ and $V^{pr}_{qs}=\langle pr|\hat{g}|qs\rangle-\langle pr|\hat{g}|sq\rangle$ with the indices $\{p,q,r,s\}$ denote atomic orbitals.
Due to the presence of two-body Coulomb (and also Breit and QED) interactions, it is not possible to obtain atomic wave functions of multi-electron systems exactly. In practice, it is dealt with by dividing the atomic Hamiltonian into an effective one-body term and residual interactions; i.e. $H=H_0 + V_{es}$. As mentioned earlier, the one-body part $H_0$ is constructed using the DHF method in this work while correlation contributions from $V_{es}$ are included at different levels of approximations through various methods that are outlined briefly below.

\subsection{CI and GASCI methods}

The DHF method is a good starting point to construct the exact atomic wave functions in a many-electron system. The complete spectrum of single-particle solutions 
obtained from the DHF procedure constitutes a one-particle basis from which one may construct determinants that approximate the wave function for
an atomic closed-shell system through an anti-symmetric Hartree-product (Slater determinant) of four-spinors
\begin{eqnarray}\label{eq:slater}
	\Phi(\mathbf{r_{1}},\mathbf{r_{2}},\cdots,\mathbf{r_{N}})=\mathbf{A}(\psi_1(\mathbf{r_{1}}),\psi_2(\mathbf{r_{1}}),\cdots,\psi_N(\mathbf{r_{N}})) .
\end{eqnarray}

In the CI method, the exact wave function $|\Psi \rangle$ of an atomic state (known as atomic state function (ASF)) is constructed by expressing it as a linear sum of all possible singly, doubly up to
$N$-tuple excited Slater determinants (referred as configuration state functions (CSFs)) with respect 
to the DHF wave function ($|\Phi_0 \rangle$). i.e.
\begin{eqnarray}
	|\Psi \rangle = C_0 |\Phi_0 \rangle + \sum_I C_I |\Phi_I \rangle + \sum_{II} |\Phi_{II} \rangle + \cdots , 
\end{eqnarray}
where $C_n$ with $n=0, I,II,\cdots$ are the CSF mixing coefficients for the respective CSFs $|\Phi_n \rangle$. In the second-quantization form, we 
can express 
as \cite{Helgaker2000}
\begin{eqnarray}\label{eq:FCI}
	| \Psi \rangle =(C_0+\sum_{ai}c^a_ia^+_a a_i+\sum_{a>b,i>j}c^{ab}_{ij}a^+_aa^+_ba_ia_j+\cdots) | \Phi_0 \rangle .
	\label{eq:FCI}
\end{eqnarray}
The second and the third terms within the brackets are the single and double excitations, respectively, expressed in terms of creation ($a^+$) and 
annihilation ($a$) operators. The coefficients of Eq. (\ref{eq:FCI}) are obtained by solving
\begin{eqnarray}
	\mathbf{H} \mathbf{C}=E_{\bf{CI}}\mathbf{C} ,
	\label{eqci}
\end{eqnarray}
where $\mathbf{H}$ is the matrix of the atomic Hamiltonian and $\mathbf{C}$ is the matrix of the expansion coefficients. The diagonalization of the Hamiltonian $\mathbf{H}$ matrix gives the spectrum of exact eigenvalues $E_{\bf{CI}}$ for the system for a given basis set. 

In practice, carrying out a full CI is impossible, so we have to choose only a small subset of determinants that carries most of the 
correlation energies. This is generally done by truncating the CI expansion. Usually, the singly and doubly excited configurations are retained 
and this truncated CI method is referred to as CISD, where `S' and `D' stand for the single and double excitations, respectively. Correlation 
energy that arises from the excitations from the single reference DHF determinant is often referred to as dynamic correlation energy. In many situations,
multi-reference determinant states are considered to take into account static correlation effects by simultaneously exciting electrons 
from all the determinants \cite{Roos2005}. In the CI approach, such a selection of reference is referred to as multi-reference CI (MRCI) method. This 
is more effective, and the diagonalization of Eq. (\ref{eqci}) can converge faster. 

A restricted active space CI (RASCI) method has been developed to account for both the dynamics and static correlations rigorously \cite{Roos1988}.
In the RASCI, the active orbital space is divided into three subspace: RAS1, RAS2, and RAS3. RAS1 is the occupied space in which at the 
most two electron holes are created; RAS3 is the unoccupied space that receives at most two electrons from RAS1 and eventually from RAS2; RAS2 is
the current active space in which all the possible excitations are considered is formed by both the occupied and unoccupied orbitals. 
Better choice of reference wave functions can be made through the GASCI method \cite{Timo2001,Timo2003,Timo2006}. The GASCI can be considered as 
the complete generalization of the RASCI method. In this method, the number of subspaces and the number of excited electrons can be arbitrary
in contrast to the restriction on the number of the subspace and the number of the excited electrons having 2 in the RASCI method. So the 
approximation made to the wave function in the CI method can be improved through the inclusion of more core-valence and core-core correlation space. 

\begin{table}[t]
	\scriptsize
	\caption{Demonstration of GASCI scheme for Ni$^{12+}$, where `min' and `max' denote the minimal and maximal number of accumulated electrons occupied in the respective core, valence, and virtual orbitals.  }	
	{\setlength{\tabcolsep}{18pt}
		\begin{tabular}{ccc ccc}\hline\hline
			GAS No.	&	min. &  max. 	&	number of orbitals	&	orbitals	&	Type	\\\hline
			I	&	2 & 4	&	2	&	$1s$,$2s$	&	core	\\
			II	&	8 &   10 	&	3	&	$2p$	&	core	\\
			III	&	10 & 12	&	1	&	$3s$	&	core	\\
			IV	&	13 & 16	&	3	&	$3p$	&	valence	\\
			V	&	16 & 16 &	150	&	$n (s,p,\cdots i)$ with $n\ge4$	&	virtual	\\ \hline\hline 
	\end{tabular}}
	\label{tab:Ni12+GAS}
\end{table}

To illustrate the application of the GASCI method, we take an example of the Ni$^{12+}$ HCI and demonstrate how the CSFs are decided. We may consider its valence configuration $3p^{4}$ with the core occupation $1s^2 2s^2 2p^6 3s^2$. The scheme of the GASCI employed for Ni$^{12+}$ is shown in Table \ref{tab:Ni12+GAS}. Here, the correlated orbitals are divided into five subsets I-V, which correspond to the core ($1s$, $2s$, $2p$, and $3s$), the valence ($3p$), and the rest (virtual) orbitals, receptively. The `minimal' (min) and `maximal' (max) number of electrons (min, max) in the I, II, and III types of space are (2, 4), (6, 8), and (10, 12), respectively. That is to say, only single and double excitations are allowed in core orbitals. When the valance orbitals $3p$ (space IV) are included, at most three electrons are allowed to excite out of the 
total core$+$valance space. According to this GASCI scheme, the possible excitations are (1) single or double excitations from the core with no 
excitation from the valence space; (2) single excitations from the core and single excitations from the valance space; (3) single excitations 
from the core and double excitations from the valance space; (4) double excitations from the core and single excitations from the valance space; 
and (5) no excitation from the core but singles, doubles and/or triple excitations from the valance space. Combining all of those possible
occupations from the core, valence and virtual orbitals forms a CI space with about $7 \times 10^9$ determinants in Ni$^{12+}$, which includes the
most dominant core-core, core-valence, and valence-valance correlations. This means that the GASCI scheme can design a complete and near-complete 
core and valence spaces based on the multi-reference configurations of CI method. 

The GASCI method is well suited to apply to the multi-valence atomic systems having open-shell configurations in the $p$, $d$, and 
$f$ shells. This method is also apt to employ to the HCIs exhibiting strong degeneracy among the energy levels due to level-crossings 
\cite{Yu2020}. The GASCI can be employed using the relativistic quantum chemistry code package DIRAC \cite{Dirac2020,Diracpaper}. 
This method has been applied to calculate atomic properties of the Sb-like Nd$^{9+}$ ion.

\subsection{FS-RCC method}

The FS-RCC method is an all-order perturbative method, and different contributions arising through the FS-RCC calculations can be understood through perturbative analysis. To understand this point, we start with the basic formulation of many-body perturbation theory to derive the expression for the ASF in the FS-RCC method. The FS-RCC method are well adaptive to the study of the single-particle or hole system, for example the B-, Al-, and Ga-like HCIs of one $p$ valence and the W$^{13+}$, Ir$^{16+}$, and Pt$^{18+}$ ions of one $4f$ hole.

The ASF $\vert \Psi_v \rangle$ of an atomic system with a closed-core and a valence orbital $v$ can be expressed by
\begin{eqnarray}
	\vert \Psi_v \rangle = \Omega_v \vert \Phi_v \rangle,
\end{eqnarray}
where $\Omega_v$ and $\vert \Phi_v \rangle$ are referred to as the wave operator and the reference state respectively. For the 
computational simplicity we choose the working reference state as the DHF wave function $\vert \Phi_c \rangle$ for the 
closed core, then the actual reference state is constructed from it as $\vert \Phi_v \rangle= a_v^{\dagger} \vert \Phi_c \rangle$ 
for the respective state with the valence orbital $v$. 

Following the form of the reference states in our approach, $\Omega_v$ can now be divided as
\begin{eqnarray}
	\Omega_v =  1+ \chi_c + \chi_v ,
\end{eqnarray}
where $\chi_c$ and $\chi_v$ are responsible for carrying out the excitations from $\vert \Phi_c \rangle$ and $\vert \Phi_v \rangle$, 
respectively, due to the residual interaction $V_r=H-H_0$ for the DHF Hamiltonian $H_0$. In a perturbative series expansion, 
we can express as
\begin{eqnarray}
	\chi_c = \sum_k \chi_c^{(k)} \ \ \text{and} \ \ \chi_v=\sum_k \chi_v^{(k)},
\end{eqnarray}
where the superscript $k$ refers to the number of times $V_r$ is considered. The $k$th order amplitudes for the $\chi_c$ and $\chi_v$ operators 
are obtained by solving the equations 
\begin{eqnarray}
	[\chi_c^{(k)},H_0]P &=& Q V_r(1+ \chi_c^{(k-1)} )P 
\end{eqnarray}
and
\begin{eqnarray}
	[\chi_v^{(k)},H_0]P &=& QV_r (1+ \chi_c^{(k-1)}+ \chi_v^{(k-1)}) P 
	- \sum_{m=1 }^{k-1}\chi_v^{(k-m)} \nonumber \\ && \times PV_r(1+\chi_c^{(m-1)}+\chi_v^{(m-1)})P, 
	\label{mbsv}
\end{eqnarray}
where the projection operators $P=\vert \Phi_c \rangle \langle \Phi_c \vert $ and $Q= 1- P$ describe the model space and the orthogonal 
space of the DHF Hamiltonian $H_0$ respectively. The energy of the state $\vert \Psi_v \rangle$ is evaluated by using an effective Hamiltonian
\begin{eqnarray}
	H_v^{eff}= P a_v H\Omega_v a_v^{\dagger} P.
	\label{efhm}
\end{eqnarray}
Using the normal order Hamiltonian $H_N= H - PHP$ in place of $H$ in the above expression, the attachment energy of a state with the 
valence orbital $v$ is evaluated. 

The above formulation is generalized to all-orders in the FS-RCC method as  
\begin{eqnarray}
	\vert \Psi_v \rangle & \equiv & \Omega_v \vert \Phi_v \rangle = e^T \{ 1+ S_v \} \vert \Phi_v \rangle
	\label{eqcc}
\end{eqnarray}
with $\chi_c= e^T-1$ and $\chi_v=e^TS_v-1$, where $T$ and $S_v$ are the CC excitation operators that excite electrons from the core and core along
with the valence orbitals to the virtual space respectively. In the singles and doubles approximation (RCCSD method), it is given as
\begin{eqnarray}
	T=T_1 +T_2 \ \ \ \text{and} \ \ \ S_v = S_{1v} + S_{2v}.
\end{eqnarray}
The amplitudes of these operators are evaluated using the equations
\begin{eqnarray}
	\langle \Phi_c^* \vert \overline{H}_N  \vert \Phi_c \rangle &=& 0
	\label{eqt}
\end{eqnarray}
and 
\begin{eqnarray}
	\langle \Phi_v^* \vert \big ( \overline{H}_N - \Delta E_v \big ) S_v \vert \Phi_v \rangle &=&  - \langle \Phi_v^* \vert \overline{H}_N \vert \Phi_v \rangle ,
	\label{eqsv}
\end{eqnarray}
where $\vert \Phi_c^* \rangle$ and $\vert \Phi_v^* \rangle$ is the excited state configurations for the DHF states $\vert \Phi_c 
\rangle$ and $\vert \Phi_v \rangle$ respectively and $\overline{H}_N= \big ( H_N e^T \big )_l$ with subscript $l$ represents for the linked terms
only. Here $\Delta E_v = H_v^{eff} - H_c^{eff}$ is the attachment energy of the electron in the valence orbital $v$ with $H_c^{eff}= P H \big 
( 1+ \chi_c \big ) P$. Following Eq. (\ref{efhm}), expression for $\Delta E_v$ is given by
\begin{eqnarray}
	\Delta E_v  = \langle \Phi_v \vert \overline{H}_N \left \{ 1+S_v \right \} \vert \Phi_v \rangle .
	\label{eqeng}
\end{eqnarray}

Contributions from important triply excited configurations can be included at the cost of RCCSD method computation by defining perturbative
operators as
\begin{eqnarray}
	T_{3}^{pert}  &=& \frac{1}{6} \sum_{abc,pqr} \frac{\big ( H_N T_2 \big )_{abc}^{pqr}}{\epsilon_a + \epsilon_b + \epsilon_c - \epsilon_p
		-\epsilon_q - \epsilon_r} ,
	\label{t3eq}
\end{eqnarray}
and
\begin{eqnarray}
	S_{3v}^{pert} &=& \frac{1}{4} \sum_{ab,pqr} \frac{\big ( H_N T_2 + H_N S_{2v} \big )_{abv}^{pqr}}{\epsilon_a + \epsilon_b + \epsilon_v - \epsilon_p -\epsilon_q - \epsilon_r} ,
	\label{s3eq}
\end{eqnarray}
where $\{a,b,c \}$ and $\{ p,q,r \}$ represent the occupied and virtual orbitals, respectively, and $\epsilon$s are their corresponding 
DHF orbital energies.

The transition matrix element and the expectation value of any operator $O$ between the fine-structure states $|\Psi_i\rangle$ and $|\Psi_f\rangle$
are calculated in terms of the expression
\begin{equation}
	\frac{\langle\Psi_f|O|\Psi_i\rangle}{\sqrt{\langle \Psi_i | \Psi_i \rangle \langle \Psi_f | \Psi_f \rangle}}=\frac{\langle\Phi_f|\tilde{O}_{fi}|\Phi_i\rangle}
	{\sqrt{\langle\Phi_i|\tilde{N}_i|\Phi_i\rangle\langle\Phi_f|\tilde{N}_f|\Phi_f\rangle}}
	\label{eq:OP}
\end{equation}
where $\tilde{O}_{fi}=\{1+S^{\dagger}_f\}e^{T^{+}}Oe^T\{1+S_i\}$ and $\tilde{N}_{k=f,i}=\{1+S^{+}_k\}e^{T^{+}}e^T\{1+S_k\}$. In
the expectation value evaluation, it turns out to be $|\Psi_i\rangle=|\Psi_f\rangle$. We adopt iterative procedures
\cite{Sahoo-PRL-2018,Sahoo-PRA-2018} to account for contributions from both the non-terminating series $e^{T^{+}}Oe^T$ and 
$e^{T^{+}}e^T$ that appear in Eq. (\ref{eq:OP}). 

\section{Systematics of HCI clocks}
One of the most important aspects of modern atomic clocks is to achieve very high-precision measurements such that they can be applied for probing fundamental physics, including the variation of $\alpha$. Thus, systematic effects observed in an experiment play essential roles in deciding whether a transition frequency measurement in an atomic system is suitable for undertaking the task. From this point of view, it is imperative to determine the significance of some of the noted systematics that needs to be analyzed in atomic systems before considering them in the experiments. The major systematics responsible for deciding an HCI clock's accuracy is Stark shifts due to lasers, BBR shifts, thermal radiation shifts, magnetic field shifts, motion-induced shifts, micromotion shifts, collisional shifts, etc., to cite a few. These systematics are also commonly seen in the neutral atom and singly charged ion-based atomic clocks. Determining each of these effects would require performing separate experiments. However, as discussed earlier, they can be estimated quite accurately by employing potential relativistic many-body methods. In fact, prior theoretical studies can also guide the experimental to decide about the conditions like the ac and dc electric field strengths, gradient, orientation, and polarization of the quantized field, etc., in an atomic clock experiment. Though the general perception is that HCIs can have small systematics on the ground that their orbitals are very much contracted, in some cases, one of the systematics can be too large due to either degeneracy of the states or other factors. Therefore, it is essential to analyze the above systematics {\it a prior} by employing reliable many-body 
methods before a HCI is undertaken in the experiment to achieve clock frequency measurement below 10$^{-19}$ level. Here we discuss the formulations of some of the aforementioned systems and how they are determined theoretically. Other systematics, such as the motional and collisional shifts, are non-trivial to estimate theoretically and depend on environmental conditions and are not discussed here. Still, they can be controlled well by utilizing currently available well-advanced ion trap techniques \cite{Berkeland1998,Rosenband2008,Keller2015,Dube2014,Huang2017}.

\subsection{Electric quadrupole shift}
One of the most important and dominating systematic shifts in an atomic clock experiment is the electric quadrupole shift caused due to the gradient of the electric field ($\mbox{\boldmath$\nabla$}{ E}$) experienced by the atomic states of the clock transition during the measurement. This can be estimated by calculating the expectation value of the corresponding interaction Hamiltonian $H_Q=-\mbox{\boldmath$\nabla$}{E} \cdot
\mbox{\boldmath$\Theta$}(\gamma,K)$ as
\begin{eqnarray}
	\Delta E_{Quad} &=& \langle\gamma K, M_K=K |H_Q|\gamma K, M_K=K \rangle,
\end{eqnarray}
where $K$ is the angular momentum of the state with its component $M_K$, $\gamma$ represents for other quantum numbers such as parity and $\mbox{\boldmath$\Theta$}(\gamma,K)$ is known as quadrupole moment, which is the expectation value of the electric quadrupole operator $\Theta=\frac{e}{2}(3z^2-r^2)$, of the state. Using the Wigner-Eckart theorem. The electric quadrupole moment can be written as
\begin{eqnarray}
	\mbox{\boldmath$\Theta$}(\gamma,K) &=& \langle \gamma K K | \Theta | \gamma K K \rangle \nonumber \\
	&=& \left ( \begin{array}{ccc}
		K & 2 & K \\
		-K & 0 & K
	\end{array} \right )  \langle \gamma K ||\mbox{\boldmath$\Theta$} || \gamma K  \rangle.
\end{eqnarray}
For the hyperfine level, $K \equiv F$, the quadrupole shift can be expressed as
\begin{equation}
	E_{Quad}(\gamma JFM_F)=-K_1K_2\Theta(\gamma,J),
	\label{eqfq}
\end{equation}
where
\begin{equation}
	K_1=\frac{2A_E[3M_F^2-F(F+1)]}{\sqrt{(2F+3)(2F+2)(2F+1)2F(2F-1)}}
\end{equation}
and
\begin{equation}\label{eq:EquadF}
	K_2=(-1)^{I+J+F}(2F+1)\Bigg\{ \begin{array}{c c c} J&2&J\\F&I&F  \end{array} \Bigg\} \Bigg( \begin{array}{c c c} J&2&J\\-J&0&J  \end{array} \Bigg)^{-1}, 
\end{equation}
for $A_E$ representing the strength of the gradient of the applied electric field. According to the angular momentum selection rules for the above expression, $\Delta E_{Quad}$ will be zero for the states with $J=1/2$ and $J=0$. The expression in Eq. \ref{eq:EquadF} ensures that quadrupole shifts in any of the hyperfine levels of the $J=1/2,0$ state is zero. However, a finite electric quadrupole shift still exists for those states of $J$ with values other than $0$ and $1/2$. In such cases, it is possible to design experimental schemes to cancel out the electric quadrupole shift by choosing appropriate hyperfine levels or averaging out the measurements. For those ions that do not have a proper combination of $F$ and $M_F$ values for the zero quadrupole shift, experimental techniques can be adopted like averaging the clock frequencies measured in the three orthogonal directions of the quantizing external field to suppress the electric quadrupole shift down to the limit below $10^{-19}$ level \cite{Itano2000,Dube2005}.

\subsection{Quadratic Stark shift}

The quadratic Stark shift of a level level $(\gamma, J, F, M_F)$ with component $M_F$ can be evaluated by
\begin{eqnarray}
	E_{Stark}(\gamma, J, F, M_F) &=& -\frac{1}{2}\alpha (\gamma, J, F) \mathcal{E}^2 \nonumber \\
	&=& -\frac{1}{2}\alpha^{(0)}(\gamma, J, F) \mathcal{E}_z^2  -\frac{1}{4}\alpha^{(2)}(\gamma, J, F) \frac{[3M_F^2-F(F+1)]}{F(2F-1)}(3 \mathcal{E}^2_z-\mathcal{E}^2),
\end{eqnarray}
where $\mathcal{E}$ and $\mathcal{E}_z$ are the total strength and strength in the z-direction of the applied electric field, $\alpha^{(0)}(\gamma, J, F)$ and $\alpha^{(2)}(\gamma, J, F)$ are the scalar and tensor components of the total polarizability $\alpha (\gamma, J, F)$ of the hyperfine level. These quantities can be related to their corresponding values in the atomic state as \cite{Itano2000}
\begin{equation}
	\alpha^{(0)}(\gamma,J,F)=\alpha^{(0)}(\gamma,J)
\end{equation}
and
\begin{eqnarray}
	\alpha^{(2)}(\gamma,J,F) &=& (-1)^{I+J+F}  \Bigg\{ \begin{array}{c c c} F&J&I\\J&F&2  \end{array} \Bigg\}  \alpha^{(2)}(\gamma, J)\nonumber \\
	&& \times \Bigg[\frac{F(2F-1)(2F+1)(2J+3)(2J+1)(J+1)}{(2F+3)(F+1)J(2J-1)}\Bigg]^{1/2} .
\end{eqnarray}
The differential values of $\alpha^{(0)}$ and $\alpha^{(2)}$ in clock transitions of HCIs are usually very small compared to the typical values obtained in the neutral atom or singly charged ion clocks. For example, for a typical value of electric field strength $E=10$ V/m, the differential Stark shift for most previously proposed HCIs can be estimated using the above relation to be far below $10^{-19}$ level.

\subsection{BBR shift}

The BBR shift of hyperfine $F$ level can be estimated using the expression
\begin{eqnarray}
	E^{E1}_{BBR} &=& -\frac{1}{2}(831.9V/m)^2\bigg[\frac{T(K)}{300}\bigg]^4 \alpha^{(0)}(\gamma,J,F) \nonumber \\
	&=& -\frac{1}{2}(831.9V/m)^2\bigg[\frac{T(K)}{300}\bigg]^4 \alpha^{(0)}(\gamma,J),
\end{eqnarray}
where $T$ in $K$ is the temperature at which the experiment is to be conducted. Using the scalar polarizabilities of the interesting clock states, the BBR shift can be estimated. It can be noticed that BBR shifts for atomic states and their hyperfine levels are the same. The environmental temperature in the HCI clock is generally far below the room temperature. In this situation, the BBR shifts are not the major limiting factors for the HCI clocks.

\subsection{Zeeman shift}

The linear Zeeman shifts in the clock transitions can be avoided by selecting $F=0$ hyperfine levels wherever possible, else $M_F=0$ 
sublevels of finite $F$ hyperfine levels. For the finite $F$ levels with non-zero $M_F$ sublevel, the linear Zeeman shift can also be removed 
technically by alternating $\pi$-polarized transitions with extreme angular momentum states with two opposite sublevels (e.g. $M_F=\pm 2$) 
during the measurements \cite{Rosenband2008}. However, it is not possible to get rid-off the second-order Zeeman shifts in such situations and
they would provide the dominant uncertainties due to the Zeeman effects. The second-order Zeeman shift of a hyperfine level $F$ with 
sublevel $M_F$ due to magnetic field strength $\mathcal{B}$ is given by \cite{Sobelman2012}
\begin{equation}
	E_{Zeem}^2=\beta_{Zeem}(\gamma, J, F,M_F) \mathcal{B}^2,
\end{equation}
where
\begin{equation}
	\beta_{Zeem}(\gamma, J, F,M_F)=-\frac{(\mu_B g_J)^2}{\hbar} \sum_{F'}\frac{|\langle F'M_F|J_z|FM_F\rangle|^2}{E_{F'}-E_F},
\end{equation}
with $J_z$ denoting the z-component of $J$, $E_F$ is the hyperfine energy level and $F'$ corresponds to all possible intermediate hyperfine
levels. The hyperfine energy level is given by \cite{Sahoo2009}
\begin{equation}
	E_F=\frac{1}{2}A_{hyf}C+B_{hyf}\frac{\frac{3}{2}C(C+1)-2I(I+1)J(J+1)}{2I(2I-1)2J(2J-1)},
\end{equation}
where $C=F(F+1)-I(I+1)-J(J+1)$. The $E_F$ values can be determined using the hyperfine structure constants $A_{hyf}$ and $B_{hyf}$. The angular 
momentum matrix element is given by
\begin{eqnarray}
	|\langle F'M_{F'}|J_z|FM_F\rangle|^2=J(J+1)(2J+1)(2F+1) 
	(2F'+1)\bigg(\begin{array}{c c c} F&1&F'\\-M_F&0&M_F  \end{array}\bigg)^2\bigg\{ \begin{array}{c c c} J&F& I \\F'&J&1 \end{array}\bigg\}^2 .
\end{eqnarray}
Thus, $\beta_{Zeem}(\gamma, J, F,M_F)$ of a hyperfine level can be determined if Land\'{e} $g_J$ factor of the atomic state is known. For a given typical 
experimental condition, $\mathcal{B}$ value can be decided suitably to estimate the second-order Zeeman shift. It is, in principle, possible
to control the strength of the magnetic field in experiments to reduce uncertainties due to the Zeeman shifts to achieve the clock frequency
measurements to the intended level of precision.

\section{Summary}

A new generation of atomic clocks that would provide ultra-precise clock frequency measurements is highly required to probe many subtle effects supporting physics beyond the Standard Model of particle physics. Atomic clocks based on highly charged ions, having a high sensitivity to variation of $\alpha$ and extremely low sensitivity to external perturbations, are understood to possess such potentials. 

In the past decade, many HCIs have been proposed as candidates for making an ultra-precise atomic clock, which has highly stable laser-accessible clock transition, enhanced sensitivity to variation of $\alpha$, and advantageous atomic properties to inhibit external perturbations. Based on two basic rules that outline the M1 and E2 forbidden transition in fine-structure splittings and the higher order forbidden transition in complex configuration accompanied with orbital crossing, a shortlist of HCI clock candidates are summarized, which includes Ar$^{13+}$, Ni$^{12+}$, Ba$^{4+}$, Pb$^{2+}, $Ir$^{17+}$, Cf$^{15-17+}$, Nd$^{13}$, Sm$^{15+}$, Pr$^{9+}$, Nd$^{9+}$, Ho$^{14+}$, etc. The first atomic clock based on Ar$^{13+}$ has been realized now and reached $10^{-17}$ frequency uncertainty at the moment. The other ions aside from Ar$^{13+}$ can offer better frequency stability and higher sensitivity to variation of $\alpha$, indicating a new possibility for making atomic clocks. These discussions would help understand the merits and dis-merits of those HCIs and also guide seeking new highly charged ions candidates for clock and applications in searching for the variation of $\alpha$.

Knowledge of spectroscopic data essential to analyze the feasibility of considering a highly charged ion for an atomic clock experiment is currently limited. In order to realize those proposed HCI clocks, more efforts are needed, including more accurate data prediction for energies that help identify clock transition lines and various atomic properties needed for a full assessment of the systematic shifts. High-accuracy relativistic many-body methods are employed to fulfil such tasks by using variants of relativistic configuration and coupled-cluster methods. The challenges are the complete treatment of the complex electron correlation often met in strongly open $4f$-shelled configurations mixed with the $5s$ and $5p$ valence electrons. Rigorous treatment contributions from QED in fine structure splitting require accurate determination of energies and transition probabilities. 

The proposed HCIs are the prospective candidates who offer the most precise frequency standards. With theoretical and experimental progress, the atomic clock based on new HCI candidates will lead to ideal platforms to probe temporal variation of $\alpha$. Upon the fruitful accomplishment of such clocks, they can promote other high-precision experiments and invoke new exciting avenues for researching fundamental physics.

\section*{Conflict of Interest Statement}

The authors declare that the research was conducted in the absence of any commercial or financial relationships that could be construed as a potential conflict of interest.

\section*{Author Contributions}

Y.Y and B.K.S contribute equally to this work and share the first authorship. B.B.S contributed to the part contents of the CI method as well as the relativistic effects and level-crossings in HCIs.

\section*{Funding}

This work is supported by The National Key Research and Development Program of China (2021YFA1402104) and the National Natural Science Foundation of China under Grant No. 11874064, and Project supported by the Space Application System of China Manned Space Program. B.K.S. would like to acknowledge use of Vikram-100 HPC of Physical Research Laboratory (PRL), Ahmedabad.

\bibliographystyle{unsrt}{Frontiers-Vancouver} 

\bibliography{test2}

\end{document}